\documentclass[11pt]{article}
\usepackage[letterpaper]{geometry}
\usepackage{amsmath}
\usepackage{graphicx,psfrag,epsf}
\usepackage{enumerate}
\usepackage{natbib}
\usepackage{url} 
\usepackage{rotating}
\usepackage{threeparttable}
\usepackage{dsfont} 

\usepackage{hyperref}
\usepackage{authblk}

\newtheorem{lem}{Lemma}
\newtheorem{prop}{Proposition}
\newtheorem{thm}{Theorem}
\newtheorem{assmp}{Assumption}

\usepackage{multirow}

\usepackage{todonotes}
\usepackage[final]{changes}
\definechangesauthor[name={AChin}, color=magenta]{AChin}
\definechangesauthor[name={HQ}, color=red]{HQ}
\definechangesauthor[name={TC}, color=blue]{TC}

\def\suppWX{\mathcal{S}_{WX}} 
\def\1{\mathds{1}}  
\def\P{\mathbf{P}} 
\def\E{\mathbf{E}} 

\title{Testing for Unobserved Heterogeneous Treatment Effects with Observational Data}
\author[a,b,c,d]{Yu-Chin Hsu}
\affil[a]{\normalsize Institute of Economics, Academia Sinica}
\affil[b]{\normalsize Department of Finance, National Central University}
\affil[d]{\normalsize Department of Economics, National Chengchi University}
\affil[e]{\normalsize CRETA, National Taiwan University}

\author[f]{Ta-Cheng Huang$^*$}
\affil[f]{\normalsize Global Asia Institute, National University of Singapore}

\author[g]{Haiqing Xu}
\affil[g]{\normalsize Department of Economics, University of Texas at Austin}
\date{\today}

\begin{document}

\def\spacingset#1{\renewcommand{\baselinestretch}%
{#1}\small\normalsize} \spacingset{1}

%
\thispagestyle{empty}
\maketitle

\begin{abstract}
Unobserved heterogeneous treatment effects have been emphasized in the recent policy evaluation literature \citep[see e.g.,][]{heckman2005structural}. This paper proposes a nonparametric test for unobserved heterogeneous treatment effects in a treatment effect model with a binary treatment assignment, allowing for individuals' self-selection to the treatment. Under the standard \textit{local average treatment effects} assumptions, i.e., the no defiers condition, we derive testable model restrictions  for the hypothesis of unobserved heterogeneous treatment effects. Also, we show that if the treatment outcomes satisfy a monotonicity assumption, these model restrictions are also sufficient. 
Then, we propose a modified Kolmogorov-Smirnov-type test which is consistent and simple to implement. Monte Carlo simulations show that our test performs well in finite samples. For illustration, we apply our test to study  heterogeneous treatment effects of the Job Training Partnership Act on earnings and the impacts of fertility on family income, where the null hypothesis of homogeneous treatment effects gets rejected in the second case but fails to be rejected in the first application. 
\end{abstract}

\noindent {\it Keywords:}  endogenous treatment assignment, local average treatment effects, nonseparable model,  unobserved heterogeneous treatment effects

\medskip

\vfill{}
\noindent \footnotesize{$^*$Author of correspondece: \href{mailto:tchuang@nus.edu.sg}{tchuang@nus.edu.sg}. \\
\noindent Acknowledgment: The authors are grateful to the co-editor Yoon-Jae Whang, three anonymous referees, Jason Abrevaya, Qi Li, Xiaojun Song and Quang Vuong for valuable comments and suggestions on previous versions of the paper. Yu-Chin Hsu gratefully acknowledges the research support from Ministry of Science and Technology of Taiwan (MOST2628-H-001-007, MOST110-2634-F-002-045), Academia Sinica Investigator Award of Academia Sinica (AS-IA-110-H01), and Center for Research in Econometric Theory and Applications (110L9002) from the Featured Areas Research Center Program within the framework of the Higher Education Sprout Project by the Ministry of Education of Taiwan. Ta-Cheng Huang is indebted to Qi Li for his continued inspiration, guidance and support. Haiqing Xu would like to dedicate this paper to the memory of Professor Halbert White. }

\newpage
\spacingset{1.2} 
\section{Introduction}
\fontsize{11}{16pt}\selectfont 
Heterogeneous treatment effects due to unobserved latent variables have been emphasized in the policy evaluation literature. See e.g., \cite{IA_1994_ECMA}, \cite{heckman1997making}, \cite{heckman2001policy, heckman2005structural}, \cite{abadie2002instrumental}, \cite{abadie2002bootstrap,abadie2003semiparametric}, \cite{blundell2003endogeneity} , \cite{matzkin2003nonparametric}, \cite{Chesher_2003_ECMA,Chesher_2005_ECMA}, \cite{CH_2005_ECMA}, \cite{florens2008identification}, \cite{imbens2009identification}, \cite{frolich2013unconditional},  \cite{d2015identification}, and  \cite{torgovitsky2015identification}, 
among many others. In the empirical study of treatment effects using observational data, the interpretation of the widely used instrumental variable (IV) estimation relies on the key assumption that after we control for covariates, treatment effects are homogeneous across individuals.  In the presence of unobserved heterogeneous  treatment effects, the standard IV approach only estimates the \textit{local average treatment effects} (LATE), rather than the \textit{average treatment effects} (ATE)\replaced[id=TC]{; see}{. See} \cite{IA_1994_ECMA} and \cite{imbens2010better}. 

In this paper, we develop a nonparametric test for the (unobserved) heterogeneous treatment effects.   We 
model the unobserved heterogeneous treatment effects by a nonparametric and nonseparable model, i.e., the error terms are not additively separable from the treatment indicator.  
Together with the endogeneity issue introduced due to individuals' self-selection to treatment, it is well known in the literature that identification and estimation of nonseparable models are challenging. 
On the other hand, the homogeneous treatment effects assumption substantially simplifies econometrics analysis of treatment effects, since it implies that  ATE is the same as LATE, after controlling for observed heterogeneity (i.e., covariates).  
For instance, \cite{angrist1991does} use a two-stage least squares approach to estimate treatment effects of compulsory schooling on earnings.  
Therefore, if ATE is the main object of research interest, by testing for heterogeneous treatment effects, our method can assess whether  the complicated nonparametric and nonseparable treatment effect model is more appropriate (than e.g., the two-stage least squares approach) for a program evaluation assignment. 


Though important, there are only a handful of papers on testing for such unobserved heterogeneity.\footnote{There exists a substantial literature for testing observed heterogeneity, i.e., whether (conditional) average treatment effects vary across different subpopulations defined by observed covariates. 
For example, see \cite{heckman1997making},  \cite{crump2008nonparametric}, \cite{chang2015nonparametric}, \cite{abrevaya2015estimating}, \cite{athey2016recursive}, \cite{hsu2017consistent}, and \cite{lee2017doubly}, among many others.} 
In the context of ideal social experiment data, i.e., lack of endogeneity, \cite{heckman1997making} develop a lower bound for the variance of heterogeneous treatment effects, thereby providing a test for whether the data are consistent with the identical treatment effects model. 
Moreover, \cite{Hoderlein2009on} discuss specification tests for endogeneity as well as unobserved heterogeneity in nonseparable triangular models. 
Recently, \cite{LW_2014_JoE} and \cite{STU_2014_ER} establish nonparametric tests for unobserved heterogeneous treatment effects under the unconfoundedness assumption. 
In particular,  \cite{LW_2014_JoE} test unobserved heterogeneity in treatments effects via testing an equivalent independence condition on observables. 
Another closely related paper is by \cite{HSU_2010_JoE}, who test the absence of self-selection on the gain to treatment in the generalized Roy model framework, allowing for unobserved heterogeneous treatment effects. 
Furthermore, our paper is also related to \cite{heckman2005structural}, who suggest an approach to test heterogeneity of the {\it marginal treatment effects} (MTE). Our test object of interest, however, focuses on whether there exists individual-level unobserved heterogeneity in treatment effects, rather than group-level (defined by a margin) unobserved heterogeneity, i.e., whether MTE varies across margins. 

Motivated by \cite{LW_2014_JoE}, we show that in the presence of endogeneity, model restrictions arising from the homogeneous treatment effects hypothesis can also be characterized by a set of independence conditions that involve LATE. These testable implications are related to important literature on testing whether the distributions of potential outcomes are affected by the treatment. 
In the LATE framework, \cite{abadie2002bootstrap} considers the null hypothesis of the equality between distribution functions of the potential outcomes among the compliers in treatment and control groups, and also first-order and second-order stochastic dominance of the two distribution functions.  \cite{lee2009nonparametric} and \cite{chang2015nonparametric} generalize \cite{abadie2002bootstrap}'s test for conditional
distributional effects by allowing for observed treatment effects heterogeneity.\footnote{See also e.g., \cite{jun2016treatment} and \cite{hsu2017consistent} for further extensions, and references therein.} 
Our test problem differs from that literature in that we investigate whether  the distribution function of the treatment group is an (unknown) constant shift from the  control group's distribution. Specifically, the equality hypothesis on the two distribution functions is a special case of our test implication.

Nonparametric tests for conditional independence restrictions have been well studied in different contexts. 
See, e.g., \cite{andrews1997conditional}, \cite{dauxois1998nonlinear}, \cite{su2007consistent, su2008nonparametric, su2014testing}, \cite{huang2010testing}, \cite{bouezmarni2014nonparametric}, \cite{hoderlein2012nonparametric}, \cite{linton2014testing}, and \cite{huang2015flexible}, among many others. 
When one tests independence restrictions of variables that are nonparametrically constructed,  
a key technical issue arises in the case, 
in particular, in which the nonparametric components are functions of continuous covariates \citep[see, e.g.,][]{LW_2014_JoE}.  
Motivated by \cite{stinchcombe1998consistent}, we modify the classic Kolmogorov-Smirnov tests by using the primitive function of CDF's. 
Such a modification is novel and plays a key role in our approach. Moreover, we establish the asymptotic properties of the proposed tests under the null and alternative hypotheses.

The remainder of this paper is organized as follows.
In Section \ref{sec:framework_motivation}, we introduce the model and derive testable model restrictions.
Section \ref{sec:condtional_indep} discusses our test statistics and their asymptotic results.
We distinguish whether the covariates include continuous variables.
In Section \ref{sec:simulations}, we conduct Monte Carlo experiments to study the finite-sample performance of the proposed test. 
Section \ref{sec:empirical} illustrates our testing approach by two empirical applications.  All proofs are collected  in the Appendix.


\section{Model and Testable Restrictions}
\label{sec:framework_motivation}
We consider the following nonseparable treatment effect model:
\begin{equation} 
\label{model}
Y=g(D,X,\epsilon)
\end{equation}
where $Y \in \mathrm{R}$ is the outcome variable, $D \in \{0,1\}$ denotes the treatment status, $X\in\mathrm R^{d_X}$ is a vector of covariates, $\epsilon$ is an unobserved random disturbance of general form (e.g., without invoking any restriction on the dimensionality of $\epsilon$), and $g$ is an unknown but smooth function defined on $\{0,1\}\times \mathcal S_{X\epsilon}$.\footnote{For a generic random vector $A$, $\mathcal{S}_A$ denotes the support of $A$.}
In particular, the treatment variable $D$ is allowed to be correlated with $\epsilon$ to allow for selection to the treatment; see, e.g., \cite{heckman1997making}.
To deal with endogeneity, we introduce a binary instrumental variable $Z \in \{0,1\}$. 
Throughout the paper, we use uppercase letters to denote random variables, and their corresponding lowercase letters to stand for realizations of random variables.

As is motivated in the seminal paper by \cite{matzkin2003nonparametric}, the non-additivity of the structural relationship $g$ in $\epsilon$ captures the idea of unobserved heterogeneous treatment effects in that the individual treatment effect, $g(1,X,\epsilon)-g(0,X,\epsilon)$, would depend on the unobserved individual heterogeneity $\epsilon$, even after controlling for covariates $X$.
Therefore, we have the following proposition.

\begin{prop}\label{prop: 1}
Suppose  \eqref{model} holds, then the  homogeneous treatment effects hypothesis, i.e., for some measurable function $\delta(\cdot):\mathcal{S}_X \mapsto \mathrm{R}$,
\begin{equation}
\label{counterfactual}
\mathcal{H}_{0}:\  g(1,X,\cdot) - g(0,X,\cdot)=\delta(X)
\end{equation} holds if and only if  the structural relationship $g$ is additively separable in $\epsilon$ (w.r.t. $D$), i.e.,
\begin{equation}
\label{additivity}
g(D,X, \epsilon) = m(D,X) + \nu(X, \epsilon),
\end{equation}where $m: \mathcal{S}_{DX} \mapsto \mathrm R$ and $\nu:\mathcal{S}_{X\epsilon} \mapsto \mathrm R$.
\end{prop}


\noindent 
Proposition \ref{prop: 1} directly follows \cite{LW_2014_JoE}. 
Note that if \eqref{additivity} holds, $\delta(x) = m(1,x)-m(0,x)$ in \eqref{counterfactual}, which is the homogenous individual treatment effects across individuals with covariates $X=x$.

The key insight in \cite{LW_2014_JoE} is that they further show \deleted[id=TC]{that}the equivalence between the additive separability hypothesis and a conditional independence restriction
on observables. In the presence of treatment endogeneity, we derive
a similar result.  Let $p(x,z) = \P (D=1|X=x,Z=z)$ be the {\it propensity score} for each $x\in\mathcal S_X$
and $z \in \{0,1\}$. 

\noindent
\begin{assmp}
\label{assmp1}
Suppose $Z \perp \!\!\!\perp \epsilon|X$. For all $x \in \mathcal{S}_X$, $\P(Z=1|X=x)$ is bounded away from zero and one, and $p(x,0) \neq p(x,1)$.
\end{assmp}

\noindent Assumption \ref{assmp1} is standard in \added[id=TC]{the} literature\replaced[id=TC]{ and}{, which} requires the instrumental variable $Z$ to be (conditionally) exogenous and relevant.
See, e.g., \cite{IA_1994_ECMA} and \cite{CH_2005_ECMA}.  Throughout, we maintain Assumption \ref{assmp1}.
Moreover, let $\mu(x,z) = \E(Y|X=x,Z=z)$. 
Under $\mathcal{H}_0$ and Assumption \ref{assmp1}, we have
\begin{align*}
\mu(x,z) &= \E \left[g(0,X,\epsilon)+\delta(X) \times D |X=x,Z=z\right]\\
&= \E\left[g(0,X,\epsilon)|X=x\right] + \delta(x)  p(x,z), \ \ \text{for }\ z=0,1.
\end{align*} 
In the above system of equations, we treat $\E\left[g(0,X,\epsilon)|X=x\right]$ and $\delta(x)$ as two unknowns.
Solving the equations, we then identify LATE $\delta(x)$ as follows: 
\begin{equation}
\label{eqn:late}
\delta(x) = \frac{\mu(x,1) - \mu(x,0)}{p(x,1)-p(x,0)}=\frac{\text{Cov}(Y,Z|X=x)}{\text{Cov}(D,Z|X=x)}.
\end{equation}
See \cite{IA_1994_ECMA} for the LATE interpretation of \eqref{eqn:late}. 
Note that $\delta(x)$ is well defined given $p(x,0)\neq p(x,1)$ under Assumption 1, and identified as well directly from the data regardless \added[id=TC]{of} the monotonicity of the selection.

Let $W \equiv Y + (1-D) \times \frac{\text{Cov}(Y,Z|X)}{\text{Cov}(D,Z|X)}$.
Under the null hypothesis $\mathcal{H}_0$, 
we have
\[
W = g(D,X,\epsilon) + (1-D) \times [g(1,X,\epsilon)-g(0,X,\epsilon)]=g(1,X,\epsilon)
\]
which implies that $W$ is conditionally independent of $Z$ given $X$ under Assumption \ref{assmp1}. 
Therefore, 
we obtain the following lemma. 

\begin{lem}\label{Lem: 2}
Suppose \eqref{model} and Assumption \ref{assmp1} hold. 
Then, $\mathcal{H}_0$ implies that $W \perp \!\!\! \perp Z | X$. 
On the other hand, if   $W \perp \!\!\! \perp Z | X$, then the observed data can be rationalized by a structure that satisfies $\mathcal{H}_0$.
\end{lem}

%

Lemma \ref{Lem: 2} shows that the conditional independence condition is all the testable restrictions of $\mathcal{H}_0$, i.e., it is sharp in the sense of Definition 1 of \cite{HLS2019}. 
Regarding the first part of Lemma \ref{Lem: 2}, 
intuitively, if treatment effects are homogeneous, 
we can estimate them by the IV method, 
and further construct potential outcomes that are independent of the instrumental variable.\footnote{Note that one could also define $W^a=Y -D\times \delta(X)$, which \replaced[id=TC]{is equal}{equals} to $g(0,X,\epsilon)$ under Assumption \ref{assmp1}.}    
Note that the conditional  independence condition in Lemma \ref{Lem: 2} can be equivalently rewritten as
\begin{multline*}
\frac{\P(Y\leq y; D=1|X,Z=1)-\P(Y\leq y;D=1|X,Z=0)}{p(X,1)-p(X,0)}\\
=\frac{\P(Y\leq y-\delta(X); D=0|X,Z=0)-\P(Y\leq y-\delta(X); D=0|X,Z=1)}{p(X,1)-p(X,0)},
\end{multline*}
provided \deleted[id=TC]{by}$p(X,1)\neq p(X,0)$ almost surely.
Under the additional monotonicity condition on the selection, both sides in the above equation can be interpreted as the conditional distribution of ``potential outcomes'' given the compliers group in \cite{IR_1997_ReStud}.

According to Lemma \ref{Lem: 2}, rejecting $W \perp \!\!\! \perp Z | X$ suggests the presence of unobserved heterogeneous treatment effects. 
It is worth pointing out, however, that the other direction of the above statement is also true under additional assumptions given below.  These additional assumptions have been widely used for obtaining identification of quantile treatment effects, and LATE in the  IV literature \citep[see, e.g.,][]{IA_1994_ECMA,CH_2005_ECMA}. 

\begin{assmp}[Single-index error term]
\label{assu: single index}
There exists a measurable function $\tilde g: \mathcal{S}_{DX}\times \mathrm R\mapsto \mathrm{R}$ and  $\nu:\mathcal{S}_{X\epsilon} \mapsto \mathrm {R}$ such that
\[
g(D,X,\epsilon) = \tilde g(D,X,\nu(X,\epsilon)).
\] Moreover, $\tilde g(d,x,\cdot)$ is strictly increasing in the scalar-valued index $\nu$.
\end{assmp}

\noindent  Assumption 2 imposes the monotonicity of the single-index error term, of which various simplified assumptions have also been made in the literature for identification and estimation of nonseparable functions.  For instance, among many others, \cite{matzkin2003nonparametric} and \cite{Chesher_2003_ECMA} assume that the structural function $g$ is strictly increasing in the scalar-valued error term $\epsilon$.  
Note that Assumption 2 holds under the null hypothesis $\mathcal H_0$ because \eqref{additivity} would hold under $\mathcal H_0$.   Assumption 2  narrows down the space of alternatives such that the  model restrictions derived in Lemma \ref{Lem: 2} are also sufficient to distinguish the null and alternative hypotheses.



\begin{assmp}[Monotone selection]
\label{assu: selection}
The selection to the treatment is given by
\begin{equation}
\label{selection}
D = \1 \left[ \theta(X, Z) - \eta \geq 0 \right],
\end{equation}
where $\theta$ is an unknown function, and $\eta \in \mathrm {R}$ is an error term satisfying $Z \perp \!\!\! \perp  (\epsilon, \eta)|X$.
\end{assmp}


\noindent \cite{IA_1994_ECMA} first introduce the monotone selection assumption, which is essentially the ``no defier'' condition. Moreover, \cite{Vytlacil_2002_ECMA}  shows that such a monotonicity condition is observationally equivalent to the weak monotonicity of \eqref{selection} in the error term $\eta$.  


For any  $x\in \mathcal{S}_X$, let $\mathcal C_x \equiv \{ \eta \in \mathrm{R}: \min\{\theta(x,0),\theta(x,1)\}<\eta \leq \max\{\theta(x,0),\theta(x,1)\}\}$. 
Note that $\mathcal C_x$ is called the ``complier group'' if $p(x,0)<p(x,1)$ \citep[see][for the concept of the ``complier group.'']{IA_1994_ECMA}

\begin{assmp}\label{assu: support}
The support of $g(d,x,\epsilon)$ given $X=x$ and the complier group $\mathcal C_x$ is equal to the support of $g(d,x,\epsilon)$ given $X=x$, i.e.,
$\mathcal{S}_{g(d,x,\epsilon) | X=x, \eta \in \mathcal C_x} = \mathcal{S}_{g(d,x,\epsilon)|X=x}$.
\end{assmp}

\noindent Assumption \ref{assu: support} is a support condition, first introduced by \cite{vuong2014counterfactual} as the effectiveness of the IV. 
It implies that $\mathcal{S}_{g(d,x,\epsilon) | X=x, \eta \in \mathcal C_x} = \mathcal{S}_{Y|D=d,X=x}$.\footnote{To see this, note that $\mathcal{S}_{g(d,x,\epsilon) | X=x, \eta \in \mathcal C_x} \subseteq \mathcal{S}_{g(d,x,\epsilon)|D=d,X=x}\subseteq \mathcal{S}_{g(d,x,\epsilon)|X=x}$.} 
Note that the distribution of $g(d,x,\epsilon)$ given $X=x$ and $\eta\in\mathcal C_x$ can be identified; see, e.g., \cite{IR_1997_ReStud}. 
Thus, Assumption 4 is testable. Specifically, for all $t\in\mathrm R$,
\begin{multline*}
\P[g(d,x,\epsilon)\leq t|X=x,\eta\in \mathcal C_x]\\
= \frac{\P(Y\leq t, D=d|X=x, Z=1)-\P(Y\leq t, D=d|X=x,Z=0)}{\P(D=d|X= x, Z=1)-\P(D=d|X=x, Z=0)},
\end{multline*}
from which we can identify the support $\mathcal S_{g(d,x,\epsilon)|X=x,\eta\in\mathcal C_x}$.

Assumption \ref{assu: support}  allows one to use the data to address questions involving counterfactuals of outcomes of the ``always-takers''  and the ``never-takers'' groups.  
It is possible to provide sufficient primitive conditions for Assumption \ref{assu: support}. 
For instance, if one assumes $\mathcal{S}_{\epsilon | X=x, \eta \in \mathcal C_x} = \mathcal{S}_{\epsilon|X=x}$, or even a stronger condition that $(\epsilon,\eta)$ has a rectangular support conditional on $X=x$, then Assumption \ref{assu: support} holds. 

\begin{thm}\label{thm:3}
Suppose that \eqref{model} and Assumptions \ref{assmp1}-\ref{assu: support} hold.
Then $\mathcal{H}_0$ holds if and only if  $W \perp \!\!\! \perp  Z|X$.
\end{thm}


\noindent
Recall the definition of $W$;Theorem \ref{thm:3} shows that testing the null hypothesis $\mathcal{H}_0$ should just rely on  the information from the compliers group.  It is worth pointing out that Theorem \ref{thm:3} is related to  \cite{LW_2014_JoE}, who show that  $\mathcal{H}_0$ holds if and only if  $Y-\E (Y|D,X) \perp \!\!\!\perp D|X$ under the unconfoundedness  condition (i.e., $D\perp \!\!\!\perp \epsilon|X$) and Assumption \ref{assu: single index}.  

It should also be noted that Assumption 2 is a crucial condition for the equivalence between the null hypothesis $\mathcal{H}_0$ and  the testable model restrictions $W\bot Z|X$.   \cite{CH_2005_ECMA} show that this assumption is observationally equivalent to the rank similarity assumption.  In the current literature, Assumptions 2 (or the rank similarity assumption) has been widely used  for identifying heterogeneous treatment effects. See e.g.\ \cite{CH_2005_ECMA} and \cite{vuong2014counterfactual}.

We also note that throughout, we maintain the validity of the instrument, i.e.,  Assumption \ref{assmp1}.  If this assumption is questionable, then our test should be more carefully interpreted as a joint test of Assumption \ref{assmp1} and the  homogeneity of treatment effects. 

\section{Consistent tests}
\label{sec:condtional_indep}
Based on Theorem \ref{thm:3}, we now propose tests for unobserved treatment effect heterogeneity via testing the conditional independence restriction.   Because $Z$ is binary, the conditional independence restriction in Theorem \ref{thm:3} is equivalent to
\[
F_{W|XZ}(\cdot|x,0)=F_{W|XZ}(\cdot|x,1), \  \forall \ x\in\mathcal S_X.
\] Note that the variable $W$ needs to be nonparametrically constructed from the data.
In the following discussion, we distinguish the cases where the covariates $X$ are continuous random variables because the continuous-covariates case is more difficult to deal with due to the nonparametric function $\delta(\cdot)$ in the construction of $W$. 


\subsection{Case 1: Discrete Covariates}
We first discuss the case where  $X$ takes only a finite number of values. 
Let $\{(Y_i,D_i,X_i, Z_i): i\leq n\}$ be a random sample of $(Y,D,X,Z)$. By Theorem \ref{thm:3}, we test $\mathcal{H}_0$ via the following model restrictions:
\[
F_{W|XZ}(\cdot\ |x,0) = F_{W|XZ}(\cdot\ |x,1), \ \forall \ x\in\mathcal S_X,
\]where $W = Y + (1-D)\times \delta(X)$ is generated from the observables.

For a generic $k$-dimensional random vector $(A_1,\cdots, A_k)$, we let $\1_{A_1\cdots A_k}(a_1,\cdots,a_k) \equiv \1(A_1=a_1, \cdots, A_k=a_k)$ and $\1(\cdot)$ be the indicator function. We estimate $\delta(X_i)$ as follows
\[
\hat{\delta}(x) = \frac{
\sum_{ i=1}^n Y_i \1_{X_iZ_i}(x,1) \times
\sum_{ i=1}^n \1_{X_i}(x)  - 
\sum_{ i=1}^n Y_i \1_{X_i}(x) \times
\sum_{ i=1}^n \1_{X_iZ_i}(x,1) }{
\sum_{ i=1}^n D_i  \1_{X_iZ_i}(x,1) \times
\sum_{ i=1}^n \1_{X_i}(x) -
\sum_{ i=1}^n D_i \1_{X_i}(x) \times 
\sum_{ i=1}^n \1_{X_iZ_i}(x,1) },
\]
and \replaced[id=TC]{further let}{let further} $\widehat W_i = Y_i + (1-D_i)\times  \hat\delta(X_i)$. 
We now define our test statistic:
\[
\widehat{\mathcal{T}}_n = \sup_{(w,x) \in \suppWX} \sqrt{n}\ \left| \widehat{F}_{{W}|XZ}(w|x,0) - \widehat{F}_{{W}|XZ}(w|x,1) \right|,
\]
where $\widehat{F}_{{W}|XZ}(w|x,z) = \frac{\sum_{i=1}^n \1(\widehat{W}_i \leq w) Z_i \1_{X_i}(x)}{\sum_{i=1}^n Z_i \1_{X_i}(x)}$.

Next, we establish the asymptotic properties of the test \replaced[id=TC]{statistic}{statistics} $\widehat{\mathcal{T}}_n $. Let 
\[
f_{WD|XZ}(w,d|x,z) \equiv f_{W|DXZ}(w|d,x,z) \times \P(D=d|X=x,Z=z),
\] 
and 
\[
\kappa(w,x)\equiv - \frac{f_{WD|XZ}(w,0|x,1)-f_{WD|XZ}(w, 0|x,0)}{p(x,1)-p(x,0)}.
\]
Note that, under Assumptions \ref{assmp1} and 3, $\kappa(w,x) \geq 0$ since it becomes the conditional density of $g(0,x,\epsilon)$ given the complier group and $X=x$.

\begin{assmp}
\label{assu: 5}
Assume that
\begin{enumerate}[(i)]
\item
$X$ is discrete and takes a finite number of values and $P(X=x,Z=z)>0$ for all $(x,z)\in\mathcal{S}_{XZ}$;

\item
$p(x,1)-p(x,0)>0$ for all $x\in\mathcal{S}_X$;  

\item
$W$ has a compact support and $\partial f_{WD|XZ}(w,0|x,z)/\partial w$ is bounded above uniformly over $(w,x,z)\in\mathcal{S}_{WXZ}$.  
\end{enumerate}
\end{assmp}

Moreover, let
\begin{align}
\label{eqn:dpsi}
\psi_{wx} &\equiv \Big[ \1(W \le w) - F_{W|XZ}(w|x,1) \Big] \times \frac{\1_{XZ}(x,1)}{\P(X=x,Z=1)} \notag\\
          &\quad - \Big[ \1(W \le w) - F_{W|XZ}(w|x,0) \Big] \times \frac{\1_{XZ}(x,0)}{\P(X=x,Z=0)} \\
\label{eqn:dphi}
\phi_{wx} &\equiv \kappa(w,x) \Big[ W - \E(W|X=x,Z=0) \Big] \times \frac{\1_{XZ}(x,1)}{\P(X=x,Z=1)} \notag \\
          &\quad -  \kappa(w,x) \Big[ W - \E(W|X=x,Z=1) \Big] \times \frac{\1_{XZ}(x,0)}{\P(X=x,Z=0)} 
\end{align}
We now derive the asymptotic behavior of the test statistic. 


\begin{thm}\label{thm:4}
Suppose that \eqref{model} and Assumptions \ref{assmp1}-\ref{assu: 5} hold.   
Then, under $\mathcal{H}_0$,
\[
\widehat{\mathcal{T}}_n  \overset{d}{\rightarrow} \sup_{(w,x) \in \suppWX} | \mathcal{Z}(w,x) |,
\]
where $\mathcal{Z}(\cdot,\cdot)$ is a mean-zero Gaussian process with a covariance kernel:
\[
\text{Cov} \left[ \mathcal{Z}(w,x), \mathcal{Z}(w',x') \right] = \mathrm E \left[ (\psi_{wx} + \phi_{wx}) (\psi_{w'x'} + \phi_{w'x'})\right], \ \forall (w,x),(w',x') \in \suppWX.
\]
Moreover, under $\mathcal{H}_1$, we have
\[
n^{-\frac{1}{2}}\widehat{\mathcal T}_n\overset{p}{\rightarrow} \sup_{(w,x) \in \suppWX} \left\vert {F}_{{W}|XZ}(w|x,0) - { F}_{{W}|XZ}(w|x,1) \right\vert >0.
\]
\end{thm}


\noindent
In Appendix, we show that the influence function for $\sqrt{n}[\widehat{F}_{{W}|XZ}(w|x,0) - \widehat{F}_{{W}|XZ}(w|x,1)-{F}_{{W}|XZ}(w|x,0) + {F}_{{W}|XZ}(w|x,1)]$ is $\psi_{wx} + \phi_{wx}$, in which $\psi_{wx}$ is the influence function when $\delta(x)$ is known, and $\phi_{wx}$ accounts for the estimation effect of estimating $\delta(x)$.
By Theorem \ref{thm:4}, our test is one-sided: reject $\mathcal{H}_0$ at significance level $\alpha$ if and only if $\widehat{\mathcal T}_n> c_{\alpha}$, where $c_\alpha$ is the $(1-\alpha)$-th quantile of $\sup_{(w,x) \in \suppWX} |\mathcal{Z}(w,x)|$.

Since the asymptotic distribution of $\sup_{(w,x) \in \suppWX} |\mathcal{Z}(w,x)|$ is non-pivotal and complicated,  
we apply the multiplier bootstrap method to approximate the entire process to construct the critical value. 
See, e.g., \cite{van1996weak},  \cite{delgado2001significance}, \cite{donald2003ecma}, 
and \cite{DonaldHsu2014}. 
Specifically, we simulate a sequence of
i.i.d.\ pseudo-random variables $\{U_i: i=1,\cdots,n\}$ that is independent of  the random sample path $\{(Y_i,X_i,D_i,Z_i):
i=1,2,\cdots\}$  with
$\mathrm E[U]=0$, $\mathrm E[U^2]=1$, and $\mathrm E[U^4] <
+\infty$. Then, we obtain the following simulated empirical
process:
\[
\widehat{\mathcal{Z}}^u(w,x) = \frac{1}{\sqrt{n}} \sum_{i=1}^n U_{i} \times ( \hat{\psi}_{wx,i} + \hat{\phi}_{wx,i}),
\]
where $\hat{\psi}_{wx,i} + \hat{\phi}_{wx,i}$ is the estimated influence function. 
Namely,
\begin{align*}
\hat{\psi}_{wx,i} &=  \left[\1(\widehat{W}_i \le w) -  \frac{\sum_{j=1}^n \1(\widehat{W}_i \le w) \1_{X_jZ_j}(x, 1)}{\sum_{j=1}^n \1_{X_jZ_j}(x,1)} \right] \times  \frac{\1_{X_iZ_i}(x,1)}{\widehat{\P}(X=x,Z=1)} \\
&\quad -  \left[\1(\widehat{W}_i \le w) -  \frac{\sum_{j=1}^n \1(\widehat{W}_i \le w) \1_{X_jZ_j}(x,0)}{\sum_{j=1}^n \1_{X_jZ_j}(x,0)} \right] \times  \frac{\1_{X_iZ_i}(z,0)}{ \widehat{\P}(X=x,Z=0)} ;\\
\hat{\phi}_{wx,i} &= \hat{\kappa}(w,x)
\left[ \widehat{W}_i - \frac{\sum_{j=1}^n \widehat{W}_j \1_{X_jZ_j}(x,0)}{\sum_{j=1}^n \1_{X_jZ_j}(x,0)}\right] \times \frac{\1_{X_iZ_i}(x,1)}{ \widehat{\P}(X=x,Z=1)} \\
&\quad - \hat{\kappa}(w,x)
\left[ \widehat{W}_i - \frac{\sum_{j=1}^n \widehat{W}_j \1_{X_jZ_j}(x,1)}{\sum_{j=1}^n \1_{X_jZ_j}(x,1)}\right] \times \frac{\1_{X_iZ_i}(x,0)}{\widehat{\P}(X=x,Z=0)},
\end{align*}
where $\widehat{\P}(X=x,Z = z)$ and $\hat{\kappa}(w,x) =
-\frac{\hat{f}_{{W}D|XZ}(w,0|x,1)-\hat{f}_{{W}D|XZ}(w,
0|x,0)}{\hat{p}(x,1) - \hat{p}(x,0)}$ are uniformly consistent nonparametric estimators for 
${\P}(X=x,Z = z)$ and ${\kappa}(w,x)$, respectively.
For a given significance level $\alpha$, the critical value
$\hat{c}_{n}(\alpha)$ is obtained as the $(1-\alpha)$-quantile of
the simulated distribution of $\sup_{(w,x) \in
\mathcal{S}_{WX}} \big| \widehat{\mathcal{Z}}^u(w,x)\big|$.

Now, we give additional conditions for the validity of the multiplier bootstrap critical value. 

\begin{assmp}  
\label{assu: multiplier}
Assume that
$\{U_i: i=1,\cdots,n\}$ is a sequence of
i.i.d.\ pseudo random variables that is independent of  the random sample path $\{(Y_i,X_i,D_i,Z_i):
i=1,2,\cdots\}$  with
$\mathrm E[U]=0$, $\mathrm E[U^2]=1$, and $\mathrm E[U^4] <\infty$.
\end{assmp}

In simulations and empirical studies, we set $U_i$'s as standard normals so that Assumption \ref{assu: multiplier} is satisfied.

\begin{assmp}  
\label{assu: multi-discrete}
Assume that for $z=0,1$,
\begin{enumerate}[(i)]
\item
$\sup_{x\in\mathcal{S}_{X}}|\widehat{\P}(X=x,Z = z)-\P(X=x,Z = z) |\stackrel{p}{\rightarrow} 0$;

\item
$\sup_{x\in\mathcal{S}_{X}}|\hat{p}(x,z)-p(x,z) |\stackrel{p}{\rightarrow} 0$;

\item
$\hat{f}_{{W}D|XZ}(w,0|x,z)$ is continuous in $w$ for all $x\in\mathcal{S}_{X}$ and \\ $\sup_{(w,x)\in\mathcal{S}_{WX}} |\hat{f}_{{W}D|XZ}(w,0|x,z)-f_{{W}D|XZ}(w,0|x,z)|\stackrel{p}{\rightarrow} 0$;

\item
$\sup_{x\in\mathcal{S}_{X}}|\hat{\delta}(x)-{\delta}(x)|\stackrel{p}{\rightarrow} 0$.

\end{enumerate}
\end{assmp}

\begin{thm}\label{thm: size-discrete}
Suppose Assumptions \ref{assmp1}-\ref{assu: multi-discrete} hold.   
Then
\begin{enumerate}[(a)]
\item
under $H_0$, $\lim_{n\rightarrow \infty}P(\widehat{\mathcal{T}}_n \geq \hat{c}_{n}(\alpha))=\alpha$;

\item
under $H_1$, $\lim_{n\rightarrow \infty}P(\widehat{\mathcal{T}}_n \geq \hat{c}_{n}(\alpha))=1$.
\end{enumerate}
\end{thm}

Theorem \ref{thm: size-discrete} shows the size and power of our test for the discrete case.  The proof of Theorem \ref{thm: size-discrete} follows standard arguments once we establish the validity of the multiplier bootstrap for the processes  in that
$\widehat{\mathcal{Z}}^u(\cdot,\cdot)\Rightarrow \mathcal{Z}(\cdot,\cdot)$ conditional
on the sample path $\{(Y_i,X_i,D_i,Z_i):
i=1,2,\cdots\}$ with probability approaching one.   Assumptions \ref{assu: multi-discrete} contains high-level conditions and we provide estimators and give low-level conditions in Appendix.  Please see the discussion after the proof of Theorem \ref{thm: size-discrete}.

By Assumption \ref{assu: 5}, $X$ is assumed to be a discrete random variable taking a finite number of values.  In this case, we literally conduct the test by sample splitting in that we test the equality of conditional distributions over all subpopulations defined by covariate value $x$.  As a result, it is straightforward to extend our test to the case of discrete random vector $X$. Therefore, we omit the details for brevity. 


\subsection{Case 2: Continuous Covariates}
We now consider the case where $X$ is a vector of continuous covariates.
To extend the empirical process argument used in the proof of Theorem \ref{thm:4} to this case,  
we propose a modified Kolmogorov-Smirnov test statistic.   
Such a modification allows us to account for the estimation effects from the generated variable $W$ which is constructed from the unknown function 
$\delta(\cdot)$ as an infinite-dimensional parameter.

Let $\lambda(t)= -t \times \mathrm 1 (t\leq 0)$ and $\Pi(w|x,z)= \mathrm E [\lambda(W-w)|X=x, Z=z]$. 
Note that $\lambda(\cdot)$ is  continuous and has a directional derivative.  
By \deleted[id=TC]{the}definition, $\Pi(\cdot |x,z)
$  is the primitive function of the $F_{W|XZ}(\cdot|x,z)$, i.e.,
\[
\frac{\partial }{\partial w} \Pi(w|x,z) = F_{W|XZ}(w|x,z).
\]  
Thus, the model restriction $W \perp \!\!\!\perp Z | X$ can be characterized as follows
\[
\frac{\partial }{\partial w}\Pi(w|x,0) =\frac{\partial }{\partial w}\Pi(w|x,1), \ \ \forall (w,x)\in\suppWX, 
\]
which holds if and only if $\Pi(w|x,0) =\Pi(w|x,1)$ for all $(w,x)\in\suppWX$.\footnote{The equivalence holds by the fact that for a continuous function $f(t)$, $f(t)=0$ for all $t\in[0,1]$ if and only if $\int_0^tf(s)ds =0$ for all $t\in[0,1]$.} 

In terms of the probability distribution of $W$ given $(X,Z) = (x,z)$, both $F_{W|XZ}(\cdot|x,z)$ and
$\Pi(\cdot|x,z)$ contain the exact same amount of information. The latter, however, allows us to
derive a test statistic and establish its limiting distribution when $W$ has to be nonparametrically generated. When covariates $X$ are continuously distributed, the generated sample $\{\widehat{W}_i: i \le n\}$ involves the nonparametric component $\hat{\delta}(X_i)$. We exploit the smoothness of $\lambda(\cdot)$ and show
that this first-stage estimation error can be further aggregated out at the
$\sqrt{n}$-rate in our test statistic defined on $\{\widehat{W}_i: i \le n\}$.
It should be noted that when covariates $X$ are discrete as discussed in the last subsection, we can also apply a similar testing procedure via testing $\Pi(\cdot|x,1) = \Pi(\cdot|x,0)$. 
Moreover, we assume that $\mathcal{S}_W$ is compact for expositional simplicity.

We denote $f_{XZ}(x,z)\equiv f_{X|Z}(x|z)\times \P(Z=z)$.
We also let $\1^*_{A}(a) \equiv \1(A \leq a)$.
For $z\in\{0,1\}$, let  $z'=1-z$ and
\[
G(w,x, z)=\E \left[\lambda (W-w)\1^*_{X}(x) \1_Z(z) f_{XZ}(X,z') \right].
\] 
Motivated by \cite{stinchcombe1998consistent}, we rewrite the above conditional expectation restrictions  by the following unconditional ones:
\begin{equation}
\label{restriction_G}
G(w,x,0)=G(w,x,1), \ \ \forall (w,x)\in\suppWX.
\end{equation}
To see the equivalence, first note that
\[
G(w,x, z)=\E \left[\lambda (W-w) \1(X\leq x) f_{X|Z}(X|z') |Z=z\right] \P(Z=0) \P(Z=1).
\]
Moreover, by the law of iterated expectation,
\[
\frac{\partial}{\partial x} \E[ \lambda(W-w) \1 (X\leq x)    f_{X|Z}(X|z')| Z=z] = \Pi(w|x,z)  f_{X|Z}(x|0)f_{X|Z}(x|1).
\]
Therefore, we obtain the conditional expectation restrictions as the
derivatives of \eqref{restriction_G}.  Note that the  estimation of
$G(w,x,z)$ avoids any denominator issues, which thereafter
simplifies our  asymptotic analysis.

For a random variable $A$ and a value $a$, let $K_{A,h}(a) \equiv K((A-a)/h)/h$ where $K$ and $h$ are a kernel function and a smoothing bandwidth, respectively.  
For an $d_X$-dimensional random vector $A$, we let $K_{A,h}(a) = \Pi_{j=1}^{d_X} K_{A_j,h}(a_j)=h^{-d_X} K((A-a)/h)$.
We estimate $\delta(X_i)$ by
\[
\hat{\delta}(X_i)=\frac{\sum_{j\neq i} Y_j Z_j K_{X_j,h}(X_i)  \sum_{j\neq i} K_{X_j,h}(X_i) -
\sum_{j\neq i} Y_j  K_{X_j,h}(X_i)   
\sum_{j\neq i} Z_j K_{X_j,h}(X_i)}{\sum_{j\neq i} D_j Z_j K_{X_j,h}(X_i)  \sum_{j\neq i} K_{X_j,h}(X_i) -\sum_{j\neq i} D_j  K_{X_j,h}(X_i) \sum_{j\neq i} Z_j K_{X_j,h}(X_i)}.
\]
Note that in this paper, we consider a kernel estimator for the nonparametric components of the test.  To avoid the boundary issue, we follow the literature to trim the support of $X$. To be specific, we will assume that the support of covariates $X$ is a Cartesian product of compact intervals, $\mathcal{S}_X=\prod_{j=1}^{d_X}[x_{\ell j}, x_{uj}]$ and $\mathcal{S}_X^\xi=\prod_{j=1}^{d_X}[x_{\ell j}+\xi, x_{uj}-\xi]$ where $\xi>0$ is a small positive number.
Moreover, let $\widehat W_i = Y_i + (1-D_i)\times  \hat\delta(X_i)$ and 
\begin{align*}
&\widehat{G}(w,x, z) = \frac{1}{n}\sum_{i=1}^n 
\1(X_i\in\mathcal{S}_X^\xi)\lambda (\widehat{W}_i-w)\mathrm \1^{*}_{X_i}(x)\1_{Z_i}(z)  \hat{f}_{XZ}(X_i,z'),\\
&\hat{f}_{XZ}(X_i,z) = \frac{1}{n } \sum_{j\neq i} K_{X_j,h}(X_i) \1_{Z_j}(z).
\end{align*}

Thus, we define our test statistic as follows:
\[
\widehat{\mathcal{T}}^{c}_{n} = \sup_{w\in\mathcal{S}_W,x\in\mathcal{S}_X^\xi}\ \sqrt{n} \left| \widehat{G}(w, x, 0) - \widehat{G}(w, x, 1) \right|.
\]
In \added[id=TC]{the} above definition, the supports of $W$ and $X$ are assumed to be known for simplicity.
In practice, this assumption can be relaxed by using consistent set estimators $\widehat{\mathcal{S}}_W$ and $\widehat{\mathcal{S}}_X$.

We show that the proposed test statistic $\widehat{\mathcal T}^c_n$ converges in distribution at the regular parametric rate under the null. The key step of our proof is to show that
\begin{equation}
\label{eq9}
\sup_{w\in\mathcal{S}_W,x\in\mathcal{S}_X^\xi}\ \left| \widehat{G} (w,x,z)- \widetilde{G}(w,x,z)\right| = o_p \left( n^{-1/2} \right).
\end{equation}
where $\widetilde G(w,x,z)=\frac{1}{n}\sum_{i=1}^n 
\1(X_i\in\mathcal{S}_X^\xi) (\widehat{W}_i-w)  \1(W_i \leq w) \1^{*}_{X_i}(x)\1_{Z_i}(z)  \hat f_{XZ}(X_i,z')$. The above result requires that the nonparametric elements in the estimation of $\hat{\delta}(\cdot)$ should converge to the corresponding true values uniformly at a rate faster than $n^{-1/4}$.

\begin{assmp} \label{assu: support-continuous}
Assume that 
\begin{enumerate}[(i)]
\item 
the support of the $d_X$-dimensional  covariates $X$ is a Cartesian product
of compact intervals, $\mathcal{S}_X=\prod_{j=1}^{d_X}[x_{\ell j}, x_{uj}]$;

\item
For $z=0,1$,   
$  \inf_{x\in\mathcal{S}_{X}} f_{X|Z}(x|z)>0$, $\sup_{x\in\mathcal S_{X}}
f_{X|Z}(x|z)<\infty$, and $\inf_{x\in\mathcal S_X}|p(x,1)-p(x,0)|>0$.

\end{enumerate}
\end{assmp}

\begin{assmp}\label{assu: continuity-cont}
For $z=0,1$,  $f_{X|Z}(x|z)$,  $p(x,z)$ and $\mu(x,z)$ are continuous in $x\in \mathcal S_{X}$.
\end{assmp}

\begin{assmp}\label{assu: bias-continuous}
For some $\iota>\frac{1}{4}$, we have $h\rightarrow 0$ and $n^\iota /\sqrt {nh^{d_X}}\rightarrow 0$ as $n\rightarrow \infty$. Moreover, the first-stage estimators satisfy \added[id=TC]{the condition} that 
for $z=0,1$
\begin{align*}
&\sup_{x\in\mathcal{S}_X^\xi}\Big|\E \Big[\frac{1}{n}\sum_{j=1}^n \1_{Z_j}(z) K_{X_j,h}(x)\Big]-f_{XZ}(x,z)\Big|= O(n^{-\iota}),\\
&\sup_{x\in\mathcal{S}_X^\xi}\Big|\E \Big[\frac{1}{n}\sum_{j=1}^n D_j \1_{Z_j}(z) K_{X_j,h}(x) \Big]-p(x,z) f_{XZ}(x,z)\Big|= O(n^{-\iota}),\\
&\sup_{x\in\mathcal{S}_X^\xi}\Big|\E\Big[\frac{1}{n}\sum_{j=1}^n Y_j \1_{Z_j}(z) K_{X_j,h}(x)\Big]-\E (Y|X=x,Z=z) f_{XZ}(x,z)\Big|= O(n^{-\iota}).
\end{align*}
\end{assmp}

\noindent
Assumptions \ref{assu: support-continuous} and \ref{assu: continuity-cont}  are standard in the nonparametric estimation literature.  
Assumption \ref{assu: bias-continuous} is a high-level condition that requires the nonparametric estimation bias \replaced[id=TC]{to diminish}{diminishes} uniformly at a rate faster than $n^{1/4}$.  
Such a condition on the bias term can be satisfied under additional primitive conditions on the kernel function and the bandwidth respectively, $K(\cdot)$ and $h$, as well as the smoothness of the underlying structural functions. See, e.g., \cite{PaganUllah1999}.

\medskip
\begin{lem}
\label{lem:5}
Suppose that Assumptions \ref{assmp1}-\ref{assu: support} and \ref{assu: support-continuous}-\ref{assu: bias-continuous} hold.
Then, \eqref{eq9} holds for $z =0,1$.
\end{lem}


\noindent By Lemma \ref{lem:5}, it suffices to establish the limiting distribution of $\widetilde G(w, x, 1)-\widetilde G(w, x, 0)$ for the asymptotic properties of our test statistics. Note that in the definition of $\widetilde G(w,x,z)$, there \replaced[id=TC]{are}{contains} no nonparametric elements estimated in the indicator function.

To establish asymptotic properties for our test, we make the
following assumption.

\begin{assmp}\label{assu: bias-strengthen} 
Assume that for $z=0,1$,
\begin{align*}
\sup_{x\in\mathcal{S}_X^{\xi}}\big|\E [\hat \delta(x)]-\delta(x)\big|=o(n^{-1/2}),~~\sup_{x\in\mathcal{S}_X^{\xi}}\big|\E[ \hat f_{XZ}(x,z)]-f_{XZ}(x,z)\big|=o(n^{-1/2}).
\end{align*} 
\end{assmp}


Assumption \ref{assu: bias-strengthen} strengthens Assumption \ref{assu: bias-continuous} by requiring the bias term in the first-stage nonparametric estimation to be smaller than $o_p(n^{-1/2})$, which can be established by using higher-order kernels \citep[see, e.g.,][]{powell1989semiparametric}.

Next, we establish the asymptotic properties of the test statistic $\widehat{\mathcal{T}}^c_n $. Let $F^*_{WD|XZ}(w,d|x,z) \equiv  F_{W|DXZ}(w|d,x,z) \times \P (D=d|X=x,Z=z)$ and
\[
\kappa^c(w,x)= - \frac{F^*_{WD|XZ}(w,0|x,1)-F^*_{WD|XZ}(w, 0|x,0)}{p(x,1)-p(x,0)}.
\]
Moreover, we define 
\begin{align}
\psi^c_{wx} &= \1(X\in\mathcal{S}_X^\xi) \Bigg\{\Big[ \lambda(W-w) - \E[\lambda(W-w)|X,Z=1] \Big] \times \frac{ \1^{*}_{X}(x)\1_{Z}(0)}{f_{XZ}(X,0)} \notag\\
&\quad -  \Big[ \lambda(W-w) - \E[\lambda(W-w)|X,Z=0] \Big] \times \frac{\1^{*}_{X}(x)\1_{Z}(1)}{f_{XZ}(X,1)}  \Bigg\}  f_{XZ}(X,0) f_{XZ}(X,1);\\
\phi^c_{wx} &= \1(X\in\mathcal{S}_X^\xi) \kappa^c(w, X) \Bigg\{ \Big[ W - \E(W|X,Z=0) \Big] \times \frac{\1^{*}_{X}(x)\1_{Z}(1)}{f_{XZ}(X,1)} \notag\\
&\quad -  \Big[ W - \E(W|X,Z=1) \Big] \times \frac{\1^{*}_{X}(x)\1_{Z}(0)}{f_{XZ}(X,0)} \Bigg\}  f_{XZ}(X,0) f_{XZ}(X,1).
\end{align}
\begin{thm}\label{thm:6}
Suppose that Assumptions \ref{assmp1}-\ref{assu: support} and \ref{assu: support-continuous}-\ref{assu: bias-strengthen} hold.
Then, under $\mathcal{H}_0$,
\begin{align*}
& \widehat{\mathcal{T}}^c_n\overset{d}{\rightarrow }\sup_{w\in\mathcal{S}_W,x\in\mathcal{S}_X^{\xi}} |\mathcal Z^c(w,x)|
\end{align*}
where  $\mathcal Z^c(\cdot,\cdot)$  is a mean-zero Gaussian process with covariance kernel
\[
\text{Cov} \left[ \mathcal{Z}^c(w,x), \mathcal{Z}^c(w',x') \right] =
\E \left[(\psi^{c}_{wx} + \phi^{c}_{wx}) (\psi^{c}_{w'x'} + \phi^{c}_{w'x'})\right], \ \forall  w,w'\in\mathcal{W},~x,x'\in\mathcal{S}_X^{\xi}.
\]
Moreover, under $\mathcal{H}_1$, we have
\[
n^{-\frac{1}{2}}\widehat{\mathcal{T}}^c_n  \overset{p}{\rightarrow} \ 
\sup_{w\in\mathcal{S}_W,x\in\mathcal{S}_X^{\xi}}\  |G(w,x,0)-G(w,x,1)|>0.
\]
\end{thm}
We also show in Appendix that the influence function for $\sqrt{n}(\widehat{G}(w, x, 0) - \widehat{G}(w, x, 1)-{G}(w, x, 0) + {G}(w, x, 1))$ is $\psi^c_{wx} + \phi^c_{wx}$, in which $\psi^c_{wx}$ is the influence function when  $\delta(x)$ and $f_{XZ}(x,z)$ are known, and $\phi^c_{wx}$ accounts for the estimation effect of $\delta(x)$ and $f_{XZ}(x,z)$.  Note that the 
$\1(X\in\mathcal{S}_X^\xi)$ term in the influence function accounts for the fact that we consider a trimmed support of $X$.

Let the simulated empirical process for the continuous case be
\[
\widehat{\mathcal{Z}}^{c,u}(w,x) = 
\frac{1}{\sqrt{n}} \sum_{i=1}^n U_{i} \times ( \hat{\psi}^c_{wx,i} + \hat{\phi}^c_{wx,i}),
\]
where $\hat{\psi}^c_{wx,i} + \hat{\phi}^c_{wx,i}$ is the estimated influence function such that
\begin{align*}
\hat\psi^c_{wx,i} &= \1(X_i\in\mathcal{S}_X^\xi) \Bigg\{ \Big[ \lambda(\widehat{W}_i-w) - \widehat{\E}[\lambda(W-w)|X_i,Z_i=1] \Big] \times \frac{\1^{*}_{X_i}(x)\1_{Z_i}(0)}{\hat{f}_{XZ}(X_i,0)}\notag\\
&\quad -  \Big[\lambda(\widehat{W}_i-w) - \widehat{\E}[\lambda(W-w)|X_i,Z_i=0] \Big] \times \frac{ \1^{*}_{X_i}(x)\1_{Z_i}(1)}{\hat{f}_{XZ}(X_i,1)} \Bigg\}  \hat{f}_{XZ}(X_i,0) \hat{f}_{XZ}(X_i,1);\\
\hat\phi^c_{wx,i} &= \1(X_i\in\mathcal{S}_X^\xi) \hat{\kappa}^c(w, X_i) \Bigg\{ \Big[ \widehat{W}_i - \widehat{\E}[W|X_i,Z=0] \Big] \times \frac{\1^{*}_{X_i}(x)\1_{Z_i}(1)}{\hat{f}_{XZ}(X_i,1)}\notag\\
&\quad -  \Big[ \widehat{W}_i - \widehat{\E}[{W}|X_i,Z_i=1] \Big] \times \frac{ \1^{*}_{X_i}(x)\1_{Z_i}(0)}{\hat{f}_{XZ}(X_i,0)} \Bigg\} \hat{f}_{XZ}(X_i,0) \hat{f}_{XZ}(X_i,1).
\end{align*}
where $\hat{f}_{XZ}(X_i,z)$, $\widehat{\E}[W|X_i, Z=z]$, $\widehat{F}_{W|DXZ}(w|0,X_i,z)$ and $\hat{p}(X_i,z)$ are uniformly consistent nonparametric estimators for 
${f}_{XZ}(X,z)$, ${\E}[W|X,Z=z]$, ${F}_{W|DXZ}(w|0,X,z)$ and ${p}(X,z)$, respectively.
For a given significance level $\alpha$, the critical value
$\hat{c}^c_{n}(\alpha)$ is obtained as the $(1-\alpha)$-quantile of
the simulated distribution of $\sup_{w\in\mathcal{S}_W,x\in\mathcal{S}_X^{\xi}} \big| \widehat{\mathcal{Z}}^{u,c}(w,x)\big|$. We would reject
$\mathcal{H}_0$ at significance level $\alpha$ when $\widehat{\mathcal{T}}^c_n> \hat{c}^{c}_{\alpha}$.

Now, we give additional conditions for the validity of the multiplier bootstrap critical value. 

\begin{assmp}  
\label{assu: multi-conti}
Assume that for $z=0,1$,
\begin{enumerate}[(i)]
\item
$\sup_{x\in\mathcal{S}^\xi_X}|\hat{f}_{XZ}(x,z)-{f}_{XZ}(x,z)|\stackrel{p}{\rightarrow} 0$;
\item
$\sup_{x\in\mathcal{S}^\xi_X}|\widehat{\E}[W|X=x]-{\E}[W|X=x]|\stackrel{p}{\rightarrow} 0$;

\item
$\sup_{x\in\mathcal{S}^\xi_X}|\hat{p}(x,z)-{p}(x,z) |\stackrel{p}{\rightarrow} 0$;

\item
$\sup_{x\in\mathcal{S}^\xi_X, w\in\mathcal{S}_W}|\widehat{F}_{W|DXZ}(w|0,x,z)-F_{W|DXZ}(w|0,x,z)| \stackrel{p}{\rightarrow} 0$, and
$\widehat{F}_{W|DXZ}(w|0,x,z)$ is non-decreasing in $w$ for all $x$ and $z$;

\item
$\sup_{x\in\mathcal{S}^\xi_X}|\hat{\delta}(x)-\delta(x)|\stackrel{p}{\rightarrow} 0$.
\end{enumerate}
\end{assmp}

\begin{thm}\label{thm: size-conti}
Suppose Assumptions \ref{assmp1}-\ref{assu: support}, \ref{assu: multiplier}, and \ref{assu: support-continuous}-\ref{assu: multi-conti}  hold.   
Then
\begin{enumerate}[(a)]
\item
under $H_0$, $\lim_{n\rightarrow \infty}P(\widehat{\mathcal{T}}^c_n \geq \hat{c}^c_{n}(\alpha))=\alpha$;

\item
under $H_1$, $\lim_{n\rightarrow \infty}P(\widehat{\mathcal{T}}^c_n \geq \hat{c}^c_{n}(\alpha))=1$.
\end{enumerate}
\end{thm}

Theorem \ref{thm: size-conti} shows the size and power of our test for the continuous case.  The proof of Theorem \ref{thm: size-conti} is similar to that of the discrete case.   Note that Assumptions \ref{assu: bias-continuous}, \ref{assu: bias-strengthen} and \ref{assu: multi-conti} are high-level conditions and we provide estimators and give low-level conditions in Appendix. Please see the discussion after the proof of Theorem \ref{thm: size-conti}.  

We can extend our test to cases where the covariate vector contains both discrete and continuous variables as in our second empirical study.  We leave the details to Appendix \ref{sec: discrete-conti}.

\subsection{Computational Issue}

When the dimension of the covariates $X$ is large, there will be a computational issue. 
For example, suppose that $X=(X_1,X_2,X_3)$ contains three continuous variables.  If we use 100 grids in $W$ and 100 grids in each element of $X$, it will result in $100^4$ points of $(w,x)$ when we calculate the test statistic and the simulated critical value.  
Therefore, the computation burden can be too heavy to be practical when the number of grids we use in each dimension is large, or the dimension of covariates is large.  We suggest conducting the test based on the test statistic calculated by all combinations of two covariates only to reduce the burden. Specifically, we calculate the test statistic based on the supremum of $100^3$ grids in $(W,X_1, X_2)$,  $(W,X_1, X_3)$ and $(W,X_2, X_3)$, respectively, so that we just need to use $3\cdot 100^3$ grids to calculate the test statistic and the simulated critical value.
The procedure is similar to the one recommended by \cite{andrews2013inference}.  


\section{Monte Carlo Simulations}
\label{sec:simulations} In this section, we investigate the finite
sample performance of our tests with a simulation study. The data
are simulated as follows:
\begin{align*}
 & Y = D + X  + [\gamma+(1-\gamma) D]\times \epsilon_1; \\
 & D = \1 \left[ \Phi(\eta) \leq 0.5 \times Z \right]; \\
 & \eta = \rho \times \epsilon + \sqrt{1-\rho^2}\times \epsilon_2.
\end{align*}
where both $\epsilon_1$ and $\epsilon_2$ conform to a uniform distribution on $[-2,2]$, $\rho = 0.7$, and $Z\sim Bernoulli(p)$ with $p=0.25,0.5$ and $0.75$ respectively.\footnote{Note that we also try different values for the correlation coefficient, and all the results are qualitatively similar. Additional Monte Carlo simulation results are available upon request.}  
For simplicity, $X,Z$ and $(\epsilon,\eta)$ are mutually independent.
Moreover, $X$ is uniformly distributed on $\{1,2,3,4\}$  and on $[0,1]$ in the discrete covariates and the continuous covariates case, respectively. Furthermore, $\gamma\in[0,1]$ describes the degree of unobserved heterogeneous treatment effects in our specification. 
In particular,  $\mathcal{H}_0$ holds if and only if $\gamma=1$. 
Intuitively, the smaller  $\gamma$ is, the more power we expect from our tests.  To investigate the size and power of our tests, we choose $\gamma\in\{1,0.75, 0.5\}$.

We consider sample size $n=1000,2000,4000$, a nominal level of $\alpha=5\%$, and $1000$ Monte Carlo repetitions. 
Given $X_i = x$, to compute the suprema of the simulated stochastic processes, we use $n/20$ grids on the support of $[\min(\widehat{W}_{i:X_i=x}),\max(\widehat{W}_{i:X_i=x})]$. 
Moreover, we use  $1000$ multiplier bootstrap samples to simulate the $p$-values. Regarding the estimation of $\kappa(w,x)$, we choose the second-order Epanechnikov kernel function with the bandwidth $h_x = c \cdot\text{std}(\widehat{W}_{i:X_i=x})\cdot n^{-1/4.5}$, and we set $c \in \{ 1.7, 2, 2.34, 2.6\}$ to study the sensitivity of the test to the bandwidth.

Table \ref{table:tab1} reports rejection probabilities of our simulations in the discrete-covariates case under the null hypothesis (i.e., $\gamma=1$) and alternative hypotheses (i.e., $\gamma=0.75,0.5$).  
From Panel A, the level of our test is fairly well behaved: It gets closer to the nominal level as the sample size increases\added[id=TC]{,} and the rejection probabilities are not sensitive to the constant $c$ for the bandwidth choice. Panels B and C show that the power of the test is reasonable. 
In particular, when $\gamma$ is closer to $1$, it is more difficult to detect such a ``local'' alternative. 
Therefore, we obtain relatively \replaced[id=TC]{low}{small} power even when \added[id=TC]{the} sample size reaches $n=2000$ in Panel B. For relatively ``small'' sample size, e.g., $n=1000$, our results show that our test performs better with a larger bandwidth choice. Moreover, when $p$ (i.e., the probability of $Z=1$) is 0.5, all the results for size and power dominate the other two cases with $p=0.25,0.75$, which is expected by our asymptotic theory.

Next, we  evaluate the performance of our tests in the case where the covariates $X$ are continuous.
To compute the suprema,  we  calculate the test statistic by using $n/20$ grid points in the support $[\min_{i=1}^n(X_i),\max_{i=1}^n(X_i)]$, as well as in the support $[Q_{\widehat{W}_i}(0.05),$ $Q_{\widehat{W}_i}(0.95)]$, where $Q_{\widehat{W}_i}(\cdot)$ is the quantile function of $\widehat{W}_i$.
In the estimation of $\delta(X_i)$ and $G(w,x)$, we choose the fourth-order Epanechnikov kernel function with the bandwidth $h_n = c \times \text{std}(X_i) \times n^{-1/3}$ to reduce the bias. 
Also, we study the sensitivity of the test to the bandwidth with $c \in \{ 1.7, 2, 2.34, 2.6\}$.
Table \ref{table:tab2}  reports the size and power properties of our test, which  are qualitatively similar to those in the discrete-covariates case.

\section{Empirical Applications}
\label{sec:empirical}
\subsection{\deleted[id=TC]{The}Effect of Job Training Program on Earnings}
We now apply our tests  to study the effects of \replaced[id=TC]{a}{the} job training program on earnings, i.e.,  the \textit{National Job Training Partnership Act} (JTPA), commissioned by the Department of Labor \added[id=TC]{of the U.S.} This program \replaced[id=TC]{funded}{began funding} training from 1983 to \added[id=TC]{the} late 1990's to increase employment and earnings for participants.
The major component of JTPA aims to support training for the economically disadvantaged. The effects of JTPA training programs on earnings have also been studied by  e.g., \cite{heckman1997making} and \cite{abadie2002instrumental} under a general framework allowing for unobserved heterogeneous treatment effects. The data \replaced[id=TC]{are}{is} publicly available at \url{ https://upjohn.org/node/952}.

Our sample consists of 11,204 observations from the JTPA, a survey  dataset from over $20,000$ adults and out-of-school youths who applied for JTPA in $16$ local areas across the country between 1987 and 1989.\footnote{JTPA services are provided at 649 sites, which might not be randomly chosen. For a given site, the applicants were randomly selected for the JTPA dataset. } 
Each participant was randomly assigned  to either a program group or a control group (1 out of 3 on average). 
Members of the program group \replaced[id=TC]{were}{are} eligible to participate \added[id=TC]{in} JTPA services, including classroom training, on-the-job training or job search assistance, and other services, while members of control group \replaced[id=TC]{were}{are} not eligible for JTPA services for 18 months. 
Following the literature \citep[see, e.g.,][]{bloom1997benefits},
we use the program eligibility as an instrumental variable for the endogenous individual participation decision.

The outcome variable is individual earnings, measured by the sum of earnings in the 30-month period following the offer.
The observed covariates include a set of dummies for \replaced[id=TC]{race, high-school graduate,}{races, high-school graduates,} and marriage,  whether the applicant worked at least 12 weeks in the 12 months preceding random assignment, and also five age-group dummies (22-24, 25-29, 30-35, 36-44, and 45-54), among others. Descriptive statistics can be found in Table \ref{table:jtpa}.
For simplicity, we group all applicants into three age categories (22-29, 30-35, and 36 and above), and pool all non-White applicants as minority applicants.

To implement the test, given $X_i=x$, we use the second-order Epanechnikov kernel and set the smoothing parameter to $2.34 \times \text{std}(\widehat{W}_{i:X_i=x}) \times n^{-1/4.5}$ when we estimate ${\kappa}(x,w)$. For the critical value, we use $10,000$ multiplier bootstrap samples and search for the suprema by using $5,000$ grid points.
The $p$-value of our test is $0.5732$.  Therefore, the null hypothesis (i.e., no unobserved heterogenous treatment effects) cannot be rejected at the $10\%$ significance level.
Our results are robust to the size of bootstrap samples, the number of grid points, and the choices of bandwidth. Note that our results are consistent with \cite{abadie2002instrumental}, who estimate quantile treatment effects under a linear specification. 
In particular, one cannot reject the null hypothesis that quantile treatment effects are invariant across different quantile levels.\footnote{
We obtain a pointwise 95\% confidence interval for each of the quantile treatment effects from Table III of \cite{abadie2002instrumental} and find that these confidence intervals overlap.
We can conclude no evidence against homogeneous treatment effects because a joint confidence band is, in general, wider than a pointwise confidence band.
}

\subsection{The Impact of Fertility on Family Income}
The second empirical illustration considers the heterogeneous impacts of children on parents' labor supply and income. Recently, \cite{frolich2013unconditional} studied the heterogeneous  effects of fertility on family income within the general LATE framework.
To deal with the endogeneity of  fertility decisions, \cite{rosenzweig1980testing}, \cite{angrist1998children},  \cite{bronars1994economic} and \cite{10.2307/146376}, among many others, suggest \replaced[id=TC]{using}{to use the} twin births as an instrumental variable.

Our data use the 1\% and 5\% Census Public Use Micro Sample (PUMS) from 1990 and 2000 censuses, consisting of 602,767 and 573,437 observations, respectively. The data \replaced[id=TC]{are}{is} publicly available at \url{https://www.census.gov/main/www/pums.html}.
Similar to \cite{frolich2013unconditional}, our sample is restricted to 21- to 35-year-old married mothers with at least one child, since we use twin birth as an instrument for fertility.
The outcome variable of interest  is the family's annual labor income.\footnote{It includes wages, salary, armed forces pay, commissions, tips, piece-rate payments, cash bonuses earned before deductions were made for taxes, bonds, pensions, union dues, etc. See \cite{frolich2013unconditional} for more details.}
The treatment variable is a dummy variable that takes the value $1$ to indicate when a mother has two or more children.
The instrumental variable is also a dummy variable and it equals $1$ if the first birth is a twin.
The covariates include mother's age, race, and educational level.
Some covariates, i.e., age and years in education, are treated as continuous variables. Summary statistics can be found in Table~\ref{table:fertility}.

Similar to the previous empirical illustration, we use the fourth-order Epanechnikov kernel and set the smoothing parameter to $2.34 \times \mbox{std}(X_i) \times n^{-1/3}$ when we estimate the influence function. 
For the critical value,  we use $5000$ bootstrapped samples and search for the suprema by using $1000$ grids for each of the \replaced[id=TC]{supports of both $W$ and $X$'s.}{support of $W$ and $X$'s.}
The bandwidths are selected in the same manner as those in the JTPA case.
The $p$-values of our tests are $0.0013$ and $0.0007$ for the 1990 and 2000 censuses, respectively.
These results suggest that the null hypothesis, i.e., homogeneous treatment effects, should be rejected at all usual significance levels.

\subsection{Extensions}
When $Z$ takes multiple values rather than being binary, one could extend our approach of
testing for unobserved heterogeneous treatment effects. Namely, let $W = Y + (1-D)\delta(X)$ where $\delta(x) = \frac{\text{Cov}(Y,Z|X=x)}{\text{Cov}(D,Z|X=x)}$. Then we test $\mathcal{H}_0$ by testing $W \perp \!\!\! \perp Z | X$. 
Since $Z$ takes more than binary values in its support, this model restriction can be equivalently written as
\[
F_{W|XZ}(\cdot|x,z) = F_{W|X}(\cdot|x),\ \forall\ (x,z) \in \mathcal{S}_{XZ}
\]
or
\[
\Pi(\cdot|x,z) = \E(\lambda(W-\cdot)|X=x), \forall\ (x,z) \in \mathcal{S}_{WXZ}
\]
depending on whether covariates $X$ or instruments $Z$ contain any continuously distributed components or not.

Such a test, however, does not exploit model restrictions arising from multiplicity of $Z$.
For instance, suppose that $\mathcal{S}_Z = \{0,1,2\}$.
Under $H_0$ and Assumption 1, we have 
\[
\frac{\E(Y|X=x,Z=z)-\E(Y|X=x,Z=z)}{p(x,z) - \E(D|X=x)} = \delta(x),\ \forall x.
\]
As a matter of fact, our test does not exploit such a model restriction.

Our analysis naturally extends the case where the treatment variable $D$ takes multiple values.
For illustration, suppose $\mathcal{S}_D = \{0,1,2\}$.
Under the homogeneous treatment effects hypothesis, denote $\delta_1(x) \equiv g(1,x,\cdot) - g(0,x,\cdot)$ and $\delta_2(x) \equiv g(2,x,\cdot) - g(0,x,\cdot)$.
For $d=1,2$, let $p_d(X,Z) = P(D=d|X,Z)$ and $W_d \equiv \delta_d(X) + Y - \sum_{d'=1}^2 \1 (D=d') \times \delta_{d'}(X)$.
Note that under $H_0$, i.e., $g(d,x,\cdot) - g(0,x,\cdot) = \delta_d(x)$, we have 
\[
W_d = g(d,X,\epsilon).
\]
By a similar argument, we test for unobserved heterogeneous treatment effects by testing $W_d \perp \!\!\! \perp Z|X$ for $d=1,2$.
To complete our analysis, it suffices to establish the identification of $\delta_d(x)$.
Note that under $\mathcal{H}_0$ and Assumption 1, we have
\[
\E(Y|X=x,Z=z) = \E[g(0,X,\epsilon)|X=x] + \delta_1(x) \times p_1(x,z) + \delta_2(x) \times p_2(x,z),\ \forall z.
\]
Therefore, $\delta_d$ is identified if $\{ ( p_1(x,z),p_2(x,z) )': z \in \mathcal{S}_{Z|X=x}\}$ has the full rank. 
Note that such a rank condition requires $\mathcal{S}_{Z|X=x}$ to contain at least three values.
\clearpage
\appendix
\small

\section{Appendix: Proofs} 
\label{sec:appendix}
\subsection{Proof of Proposition \ref{prop: 1}}
\textbf{Proof:}
For the ``if'' part, under \eqref{additivity}, we have
\[
g(1,x,\epsilon) - g(0,x,\epsilon) = m(1,x) - m(0,x) \equiv \delta(x), \ \ \forall x \in \mathcal{S}_X.
\]
For the ``only if'' part, \eqref{counterfactual} implies
\[
g(d,x,\epsilon) =  d \times [ g(1,x,\epsilon)  - g(0,x,\epsilon) ] + g(0,x,\epsilon) =  d \times \delta(x) + g(0,x,\epsilon).
\]
Therefore, \eqref{additivity} holds in the sense $m(d,x) = d \times \delta(x)$ and $\nu(x,\epsilon) = g(0,x,\epsilon)$.

\subsection{Proof of Lemma \ref{Lem: 2}}
\textbf{Proof:}
The first part of Lemma \ref{Lem: 2} is straightforward given the discussion before Lemma \ref{Lem: 2}. We now show the second part. It suffices to construct a structure that can rationalize the data and also satisfy (1), Assumption 1 and $\mathcal H_0$.  Given the observed data, denoted by $F^*_{YDXZ}$, we now construct a data generating structure for it. In the following proof, we use $Q^*_{W|X}$ to denote the quantile function of $W$ given $X$, obtained from $F^*_{YDXZ}$. Similarly, we define $\delta^*(x)$ and $p^*(x,z)$. To begin with our construction, let $\epsilon\sim U[0,1]$,  $X\sim F^*_X$ and $Z\sim F^*_Z$. Moreover, let $X$, $\epsilon$ and $Z$ be mutually independent for $F_{XZ\epsilon}$.   To complete our construction, it suffices to define the probability distribution $\P(D=1|X,Z,\epsilon)$ and the function $g$ for $Y$. Let
\[
g(d,x,\tau)=Q^*_{W|X}(\tau|x)-(1-d)\times \delta^*(x)
\] and $Y=g(D,X,\epsilon)$. 
Regarding $\P(D=1|X,Z,\epsilon)$, given \added[id=TC]{that} we have constructed $F_{\epsilon|XZ}$, we can equivalently define the joint distribution of $(D,\epsilon)$ given $X$ and $Z$. Let further
\[
\P(D=1;\epsilon\leq \tau |X=x,Z=z)=p^*(x,z)\times F^*_{Y|DXZ}( Q^*_{W|X}(\tau|x)|1,x,z), \ \ \forall \ x,z.
\]

By construction, Assumption 1 and $\mathcal H_0$ are satisfied.  Thus, it suffices to show the \replaced[id=TC]{observational}{observationally} equivalence. First, let $\tau=1$ in the construction of $P(D=1;\epsilon\leq \tau |X, Z)$, \added[id=TC]{then} it follows that  $\P(D=1|X=x,Z=z)=p^*(x,z)$. Moreover, note that
\begin{multline*}
\P(Y\leq y|D=1,X=x,Z=z)= \frac{\P(g(1,x,\epsilon)\leq y; D=1|X=x,Z=z)}{ \P(D=1|X=x,Z=z)}\\
 = \frac{\P(Q^*_{W|X}(\epsilon|x)\leq y; D=1|X=x,Z=z)}{ \P(D=1|X=x,Z=z)}=F^*_{Y|DXZ}(y|1,x,z).
\end{multline*}

\subsection{Proof of Theorem \ref{thm:3}}
\textbf{Proof:}
Because Proposition \ref{prop: 1} provides the \replaced[id=TC]{``only if''}{only if} part,  then it suffices to show the \replaced[id=TC]{``if''}{if} part. 
 Suppose $W \perp \!\!\!\perp Z | X$. Let $\tilde\delta(X)=\frac{\text{Cov}(Y,Z|X)}{\text{Cov}(D,Z|X)}$. By the definition of $W$, 
 \[
 W \equiv Y + (1-D) \times \frac{\text{Cov}(Y,Z|X)}{\text{Cov}(D,Z|X)}=Y+(1-D)\tilde\delta(X).
 \]Thus, for any $y\in\mathrm R$,
\begin{multline*}
\P(Y\le y, D=1 | X, Z = 1 ) + \P (Y + \tilde\delta(X) \le y, D=0 | X, Z = 1 ) \\
=\P (Y\le y, D=1 | X, Z = 0 ) + \P (Y + \tilde\delta(X) \le y, D=0 | X, Z = 0 ).
\end{multline*}
It follows that
\begin{multline}
\label{eqn:a1}
\P(Y \le y, D=1 | X, Z = 1 ) - \P (Y \le y, D=1 | X, Z = 0 )\\
=\P (Y \le y-\tilde\delta(X), D=0 | X, Z = 0 ) - \P (Y \le y-\tilde\delta(X), D=0 | X, Z = 1 ).
\end{multline}
Denote 
\begin{align*}
\Delta_0(\tau,x) &\equiv \P (\nu(X,\epsilon) \le \tau, D=0 | X = x, Z = 1 ) - \P (\nu(X,\epsilon) \le \tau, D=0 | X = x, Z = 0 ); \\
\Delta_1(\tau,x) &\equiv \P (\nu(X,\epsilon) \le \tau, D=1 | X = x, Z = 0 ) - \P (\nu(X,\epsilon) \le \tau, D=1 | X = x, Z = 1 ).
\end{align*}
By Assumptions 1 and 3, we have
\[
\Delta_0(\tau,x) = \P (\nu(X,\epsilon) \le \tau, \eta \in \mathcal{C}_x | X = x) = \Delta_1(\tau,x)
\]
which is strictly monotone in $\tau \in \mathcal S_{\nu(X,\epsilon)|X=x,\ \eta \in \mathcal{C}_x}$. Moreover, there is $\mathcal S_{\nu(X,\epsilon)|X=x,\ \eta \in \mathcal{C}_x}= \mathcal S_{\nu(X,\epsilon)|X=x}$ under Assumptions 2 and 4.

Therefore, we have
\begin{align*}
&\P(Y \leq y, D=1 | X = x, Z = 0 ) - \P (Y \leq y, D=1 | X = x, Z = 1 )\\
&=\Delta_1(\tilde{g}^{-1}(1,x,y), x)\\
&=\Delta_0(\tilde{g}^{-1}(1,x,y), x)\\
&=\P (Y \le \tilde{g}(0,x,\tilde{g}^{-1}(1,x,y)), D=0 | X = x, Z = 1 ) \\
&\quad- \P (Y \le \tilde{g}(0,x,\tilde{g}^{-1}(1,x,y)), D=0 | X = x, Z = 0 ),
\end{align*}
where $\tilde{g}^{-1}(1,x,\cdot )$ is the inverse function of $\tilde{g}(1,x, \cdot)$ and $\tilde{g}$ is a monotone function introduced in Assumption 2.
Note that both sides are strictly monotone in $y\in \mathcal S_{\tilde{g}(1,X,V)|X=x}$ since $\Delta_d(\cdot,x)$ is strictly monotone on $\mathcal S_{\nu(X,\epsilon)|X=x}$ under Assumption 4.

Combine the above result with \eqref{eqn:a1}, then we have
\[
\tilde{g}(0,x,\tilde{g}^{-1}(1,x,y))= y-\tilde \delta(x), \ \ \ \forall x\in\mathcal S_X, \ y\in  \mathcal S_{\tilde{g}(1,x,\nu(X,\epsilon))|X=x}.
\]
Let $y= \tilde{g}(1,x,\tau)$ for some $\tau \in \mathcal{S}_{\nu(X,\epsilon)|X=x}$. 
Then the above equation becomes
\[
\tilde{g}(0,x,\tau) = \tilde{g}(1,x,\tau) -\tilde\delta(x).
\]

\subsection{Proof of Theorem \ref{thm:4}}
\textbf{Proof:}
Let $\1^*_{WXZ}(w,x,z) = \1 (W \leq w) \times \1_{XZ}(x,z)$ and $\1^*_{\widehat WXZ}(w,x,z) = \1(\widehat W \leq w) \times \1_{XZ}(x,z)$. 
Let further $\1^*_{W(\tilde \delta)XZ} (w,x,z)=\1 (W(\tilde\delta) \leq w) \times \1_{XZ}(x,z)$, where $W(\tilde \delta)=Y+(1-D)\tilde\delta(X)$, be a function indexed by $\tilde\delta(\cdot)\in\mathrm R^{\mathcal S_X}$.
By the definition, $\1^*_{W(\delta)XZ}(w,x,z)=\1^*_{WXZ}(w,x,z)$  and $\1^*_{W(\hat \delta)XZ}(w,x,z)=\1^*_{\widehat WXZ}(w,x,z)$.

We first derive the asymptotic of $\sqrt n[\widehat F_{W|XZ}(w|x,z)-F_{W|XZ}(w|x,z)]$. 
By the definition,
\[
F_{W|XZ}(w|x,z)=\frac{\E [\1^*_{WXZ} (w,x,z)]}{\E [\1_{XZ}(x,z)]}\ \text{ and }\ \widehat{F}_{W|XZ}(w|x,z) = \frac{\E_n [\1^*_{\widehat WXZ}(w,x,z)]}{\E_n [\1_{XZ}(x,z)]}.
\]
In the expectation $\E [ \1^*_{W(\hat{\delta})XZ} (\cdot, x, z )]$ discussed below, we treat $\hat \delta$ as an index rather than a random object. 
Note that \begin{multline*}
\E_n [\1^*_{\widehat WXZ} (\cdot, x, z)]
= \E_n [ \1^*_{WXZ} (\cdot, x, z)] - \E [\1^*_{WXZ} (\cdot, x, z)]+\E [ \1^*_{W(\hat{\delta})XZ} (\cdot, x, z )]\\
+ \Big\{ \E_n [\1^*_{W(\hat{\delta})XZ} (\cdot, x, z)] - \E [\1^*_{W(\hat{\delta})XZ}(\cdot, x, z)]
- \E_n [ \1^*_{W(\delta)XZ} (\cdot, x, z)] + \E [\1^*_{W(\delta)XZ} (\cdot, x, z) ] \Big\} \\
=\E_n [ \1^*_{WXZ} (\cdot, x, z)] - \E [\1^*_{WXZ} (\cdot, x, z)]+ \E [ \1^*_{W(\hat{\delta})XZ} (\cdot, x, z)]+o_p(n^{-1/2}),
\end{multline*}
where the last step follows from the fact that
$\sqrt{n}\big(\E_n [ \1^*_{W(\delta)XZ} (\cdot, x, z)]
+ \E [\1^*_{W(\delta)XZ} (\cdot, x, z) ]\big)$ is
stochastically equicontinuous by the empirical process theory \citep[see,
e.g.,][]{van2007empirical}. By Taylor expansion,
\[
\sqrt n\ \left\{\E [ \1^*_{W(\hat{\delta})XZ} (\cdot, x, z)] -F_{W|XZ}(w|x,z) \right\}
= \frac{\partial \E [\1^*_{W(\delta)XZ}(w, x, z)]}{\partial {\delta}} \times \sqrt{n} (\hat{\delta} - \delta)+o_p(1).
\]
Note that $\frac{\partial \E [ \1^*_{W(\delta)XZ} (w, x, z)]}{\partial {\delta}(x')}=0$ for all  $x'\neq x$
and  $\frac{\partial \E [\1^*_{W(\delta)XZ} (w, x, z)]}{\partial {\delta}(x)} = -f_{W|DXZ}(w|0,x,z) \times \P (D=0,X=x,Z=z)$. 
Therefore, we have
\begin{multline*}
\sqrt n\ \left\{\E [ \1^*_{W(\hat{\delta})XZ} (\cdot, x, z)] -F_{W|XZ}(w|x,z) \right\}\\
 +\sqrt n \left\{\E_n [ \1^*_{WXZ} (\cdot, x, z)] - \E [\1^*_{WXZ} (\cdot, x, z)]\right\}
 -f_{WDXZ}(w,0,x,z) \times \sqrt{n} [\hat{\delta}(x) - \delta(x)]+o_p(1).
\end{multline*}
Moreover, $\E_n [\1_{XZ}(x,z)]=\P (X=x,Z=z) +O_p(n^{-1/2})$ under the central limit theorem. 
Thus, by Slutsky's theorem, we have
\begin{align*}
&\sqrt{n} \left[\widehat F_{W|XZ}(w|x,1)-\widehat F_{W|XZ}(w |x,0)\right]-\sqrt{n} \left[ F_{W|XZ}(w|x,1)- F_{W|XZ}(w |x,0)\right]\\
&= \frac{\sqrt{n} \big\{\E_n [\1^*_{WXZ} (w,x,1)] - \E [\1^*_{WXZ} (w,x,1)] \big\}  - f_{WDXZ}(w,0,x,1) \times  \sqrt{n} [\hat{\delta}(x)-\delta(x)]}{\mathrm P(X=x,Z=1)} \\
&\quad-\frac{\sqrt{n} \big\{\E_n [ \1^*_{WXZ} (w,x,0)]  - \E [\1^*_{WXZ}(w,x,0)] \big\} - f_{WDXZ}(w,0,x,0) \times  \sqrt{n} [\hat{\delta}(x)-\delta(x)]}{\P(X=x,Z=0)} \\
&\quad+\frac{\sqrt{n} \P(W \leq w, X = x, Z = 1)}{\E_n \1_{XZ}(x,1)} -\frac{\sqrt{n} \P(W \leq w, X = x, Z = 0)}{\E_n \1 _{XZ}(x,0)} +o_p(1).
\end{align*}
By applying the Taylor expansion,  we have
\begin{multline*}
\frac{\sqrt{n} \P(W \leq w, X = x, Z = z)}{\E_n \1_{XZ}(x,z)}- \sqrt{n}\ F_{W|XZ}(w|x,z)\\
=-F_{W|XZ}(w|x,z) \times \frac{\sqrt n  \big[\E_n \1_{XZ}(x,z) -\P(X=x,Z=z)\big]}{\P(X=x,Z=z)} +o_p(1).
\end{multline*}
Moreover, applying Lemma \ref{lem:7}, 
we have
\begin{align*}
&\sqrt{n} \left[\widehat F_{W|XZ}(w|x,1) - \widehat F_{W|XZ}(w |x,0)\right]-\sqrt{n} \left[ F_{W|XZ}(w|x,1)- F_{W|XZ}(w |x,0)\right]\\
&=\sqrt{n} \E_n \left\{[ \1(W \le w) - F_{W|XZ}(w|x,1)] \times \frac{\1_{XZ}(x,1)}{ \P (X=x,Z=1)}  \right\}\\
&\quad-\sqrt{n} \E_n \left\{[ \1 (W \le w) - F_{W|XZ}(w|x,0)] \times \frac{\1_{XZ}(x,0)}{ \P (X=x,Z=0)}  \right\}\\
&\quad+\kappa(w,x) \times \sqrt n \E_n \left\{ \big[ W - \E (W|X=x,Z=0) \big] \times \frac{\1_{XZ}(x,1)}{\P(X=x,Z=1)} \right\} \\
&\quad- \kappa(w,x) \times \sqrt n \E_n \left\{ \big[ W - \E (W|X=x,Z=1) \big] \times \frac{\1_{XZ}(x,0)}{\P(X=x,Z=0)} \right\} + o_p(1).
\end{align*}

Under the null hypothesis, there is
\begin{align*}
&\sqrt n \left[\widehat F_{W|XZ}(w|x,1)-\widehat F_{W|XZ}(w |x,0)\right]\\
&=\sqrt{n} \E_n \left\{[ \1 (W \le w) - F_{W|X}(w|x)] \times \Big[\frac{\1_{XZ}(x,1)}{ \P (X=x,Z=1)}-\frac{\1_{XZ}(x,0)}{ \P (X=x,Z=0)}\Big]  \right\}\\
&\quad+\kappa(w,x) \times \sqrt n  \E_n \left\{\big[W -\E (W|X=x)\big]\times\Big[\frac{\1_{XZ}(x,1)}{\P(X=x,Z=1)}-\frac{\1_{XZ}(x,0)}{\P(X=x,Z=0)}\Big]\right\}+ o_p(1)\\
&= \frac{1}{\sqrt{n}} \sum_{i=1}^n \left( \psi_{wx,i} + \phi_{wx,i} \right)+o_p(1)
\end{align*}
where $\psi_{wx,i}$ and $\phi_{wx,i}$ are defined by
\eqref{eqn:dpsi} and \eqref{eqn:dphi}. 
Note that for each $x\in\mathcal{X}$, $\{\1 (W \le w) - F_{W|X}(w|x): w\in\mathcal{W} \}$ is a type I class of functions according to \cite{andrews1994empirical} and this implies that
$\{\1 (W \le w) - F_{W|X}(w|x): w\in\mathcal{W}, x\in\mathcal{X} \}$ satisfies Pollard's entropy condition by Theorem 3 of \cite{andrews1994empirical}.  Note that $\frac{\1_{XZ}(x,1)}{ \P (X=x,Z=1)}-\frac{\1_{XZ}(x,0)}{ \P (X=x,Z=0)}$ is a measurable function, so it follows that 
\begin{align*}
\left\{\psi_{wx}=(\1 (W \le w) - F_{W|X}(w|x)) \cdot \left(\frac{\1_{XZ}(x,1)}{ \P (X=x,Z=1)}-\frac{\1_{XZ}(x,0)}{ \P (X=x,Z=0)}\right) : w\in\mathcal{W}, x\in\mathcal{X} \right\}
\end{align*} satisfies Pollard's entropy condition.   
Also, by Assumption \ref{assu: 5}, we have that for all $x$, 
\begin{align*}
\left\{\phi_{wx}=\kappa(w,x)(W-E[W|X=x]) \cdot \left(\frac{\1_{XZ}(x,1)}{ \P (X=x,Z=1)}-\frac{\1_{XZ}(x,0)}{ \P (X=x,Z=0)}\right) : w\in\mathcal{W}\right\}
\end{align*}
is a VC class of function according to \cite{kosorok2008introduction} and it follows that 
\begin{align*}
\left\{\phi_{wx}=\kappa(w,x)(W-E[W|X=x]) \cdot \left(\frac{\1_{XZ}(x,1)}{ \P (X=x,Z=1)}-\frac{\1_{XZ}(x,0)}{ \P (X=x,Z=0)}\right) : w\in\mathcal{W}, x\in\mathcal{X} \right\}
\end{align*}
is a VC class of function that satisfies Pollard's entropy condition.  It follows that $\{\psi_{wx}+\phi_{wx}: w\in\mathcal{W}, x\in\mathcal{X}\}$ satisfies Pollard's entropy condition.  Then combined with the previous results and by Theorem 1 of  \cite{andrews1994empirical}, we can show that 
\begin{align}\label{eq: process result}
 \sqrt{n} ( \widehat{F}_{\widehat{W}|XZ}(w|x,0) - \widehat{F}_{\widehat{W}|XZ}(w|x,1) 
 -({F}_{{W}|XZ}(w|x,0) - {F}_{{W}|XZ}(w|x,1)))\Rightarrow  \mathcal{Z}(w,x).
\end{align}
Then by the continuous mapping theorem  \citep[see, e.g.,][]{van2007empirical}, and under null,  
we have $\widehat{\mathcal T}_n \overset{d}{\rightarrow} \ \sup_{w \in \mathrm{R};  \ x\in\mathcal S_X}\  |\mathcal{Z}(w,x)|$.  In addition, (\ref{eq: process result}) implies that
\begin{align*}
 \sup_{(w,x)\in S_{WX}} \Big|( \widehat{F}_{\widehat{W}|XZ}(w|x,0) - \widehat{F}_{\widehat{W}|XZ}(w|x,1) 
 -({F}_{{W}|XZ}(w|x,0) - {F}_{{W}|XZ}(w|x,1)))\Big|\stackrel{p}{\rightarrow}
 0.
\end{align*}
It follows that under $H_1$, we have $n^{-1/2}\widehat{\mathcal{T}}_n\stackrel{p}{\rightarrow}
\sup_{(w,x) \in  S_{WX}} |F_{W|XZ}(w|x,0) - F_{W|XZ}(w|x,1)|>0$.

\subsection{Proof of Theorem \ref{thm: size-discrete}}
The proof of Theorem \ref{thm: size-discrete} would follow standard arguments once we establish the validity of the multiplier bootstrapped processes.  Note that under Assumption \ref{assu: multi-discrete} we have that
\begin{align*}
\Big|\frac{1}{n}\sum_{i=1}^n (\hat{\psi}_{wx,i}+\hat{\phi}_{wx,i})\cdot (\hat{\psi}_{w'x',i}+\hat{\phi}_{w'x',i})
-Cov[\mathcal{Z}(w,x), \mathcal{Z}(w',x')]\Big|\stackrel{p}{\rightarrow}0
\end{align*}
uniformly over $((w,x),(w',x'))\in\mathcal{S}_{WX}^2$ by the uniform law of large numbers and the uniform consistency of the various estimators in the estimated influence functions.  That is, the covariance kernel of the simulated processes converges to the covariance kernel of $\mathcal{Z}(w,x)$ uniformly.  Then by similar arguments as in \cite{hsu2017consistent}, we can show that $\mathcal{Z}^u(\cdot,\cdot)\stackrel{p}{\Rightarrow}\mathcal{Z}(\cdot,\cdot)$.  Given this result, the size and power properties of our test follow standard arguments such as \cite{andrews1997conditional}. 

\noindent{\bf Discussion on Assumption \ref{assu: multi-discrete}:}   
Here we provide estimators and low-level conditions so that Assumption \ref{assu: multi-discrete} would be satisfied.  It is straightforward to see that $\hat{\delta}(x)$ satisfies Assumption \ref{assu: multi-discrete} (iv). For $x,z\in\mathcal{S}_{XZ}$, let
\begin{align}
&\widehat{\P}(X=x,Z=z)=\frac{1}{n}\sum_{i=1}^n \1_{XZ}(x,z),\label{eq: P estimators}\\
&\hat{p}(x,z)=\frac{1}{n}\sum_{i=1}^n \1_{DXZ}(1,x,z)/\frac{1}{n}\sum_{i=1}^n \1_{XZ}(x,z)\nonumber.
\end{align}
Because $X$ and $Z$ take on a finite number of values, it is straightforward to see that $\widehat{\P}(X=x,Z=z)$ and $\hat{p}(x,z)$ given in (\ref{eq: P estimators}) satisfy Assumption \ref{assu: multi-discrete} (i) and (ii), respectively. 

We propose an estimator for $f_{WD|XZ}(w,0|x,z)$.  For $w\in\mathcal{W}=[w_\ell,w_u]$, let 
\begin{align*}
\tilde{f}_{WD|XZ}(w,0|x,z)=
\frac{\sum_{i=1}^n K_{W_i,h}(w) \1(D_i=0) \1_{X_iZ_i}(x,z)}{\sum_{i=1}^n \1_{X_iZ_i}(x,z)}
\end{align*}
and 
\begin{align}\label{eq: f_WD|XZ estimator}
\hat{f}_{WD|XZ}(w,0|x,z)=\left\{ 
\begin{array}{ll}
\tilde{f}_{WD|XZ}(w_\ell+h,0|x,z)& \text{if $w\in[w_\ell,w_\ell+h)$;}\\
\tilde{f}_{WD|XZ}(w,0|x,z)& \text{if $w\in(w_\ell+h,w_u-h)$;}\\
\tilde{f}_{WD|XZ}(w_u-h,0|x,z)& \text{if $w\in(w_u-h,w_u]$.}
\end{array}\right.
\end{align}

\begin{assmp}\label{assu: f_WD|XZ}
Assume that
\begin{enumerate}[(i)]
\item
$K(u)$ is non-negative and has support
$[-1, 1]$. $K(u)$ is symmetric around 0 and is continuously differentiable
of order 1.

\item
The bandwidth $h$ satisfies $h\rightarrow 0$, $nh^4\rightarrow\infty$ and 
$nh/\log(n)\rightarrow \infty$ as $n\rightarrow \infty$. 
\end{enumerate}
\end{assmp}

Then under Assumption \ref{assu: f_WD|XZ} and by the argument of the proof of  Lemma 3.2 of \cite{donald2012incorporating}, we show that the $\hat{f}_{WD|XZ}(w,0|x,z)$ given in (\ref{eq: f_WD|XZ estimator}) satisfies Assumption \ref{assu: multi-discrete}. Finally, for kernel functions, we can pick an Epanechnikov kernel that would satisfy Assumption \ref{assu: f_WD|XZ} (i) and for $h$, we can set $h=\hat{\sigma}_w \cdot n^{-1/4.5}$.  

\subsection{Proof of Lemma \ref{lem:5}}
\textbf{Proof:}
Fix the value of $X$ as $x$, and let $z=1$ without loss of generality.
Note that
\begin{align*}
&\widehat G(w,x,1)-\widetilde G(w,x,1) \\
&=  \E_n \left\{ \1^{*}_{XZ}(x,1)  \hat{f}_{XZ}(X,0)(w-\widehat W) \left[ \1(\widehat W \leq w) - \1( W \leq w)\right] \right\} \\
&= \E_n \Big\{ \1^{*}_{XZ}(x,1) \hat{f}_{XZ}(X,0)(w-\widehat W) \left[ \1(\widehat W \leq w) - \1( W \leq w) \right] \times \1 (|W-w|\leq  n^{-r}) \Big\}\\
&\quad+ \E_n \Big\{ \1^{*}_{XZ}(x,1)  \hat{f}_{XZ}(X,0)(w-\widehat W)  \left[ \1(\widehat W \leq w)-\1( W \leq w) \right] \times  \1 (|W-w|>  n^{-r}) \Big\}\\
&\equiv \mathrm T_1+\mathrm T_2
\end{align*}where $r\in(\frac{1}{4}, \iota)$. It suffices to show both $\mathrm T_1$ and $\mathrm T_2$ are $o_p(n^{-\frac{1}{2}})$.

First, note that
\begin{multline*}
\mathrm T_1= \E_n\Big\{\1^{*}_{XZ}(x,1) \hat{f}_{XZ}(X, 0)  (w- W)  \left[ \1(\widehat W \leq w)- \1( W \leq w) \right] \times  \1 (|W-w|\leq  n^{-r})\Big\}\\
+\E_n \Big\{\1^{*}_{XZ}(x,1) \hat{f}_{XZ}(X,0)( W- \widehat W)\left[ \1(\widehat W \leq w)-\1( W \leq w)\right] \times  \1 (|W-w|\leq  n^{-r})\Big\}.
\end{multline*}
Because
\begin{multline*}
\E \left\{ \left|\1^{*}_{XZ}(x,1) \hat{f}_{XZ}(X, 0)  (w- W)  \left[ \1(\widehat W \leq w)-\1( W\leq w) \right] \times  \1 (|W-w|\leq  n^{-r})\right| \right\} \\
\leq \E \left\{ \left| \hat{f}_{XZ}(X_1,0)\times (w- W)\times  \1 (|W-w|\leq  n^{-r})\right| \right\} =O(1)\times O(n^{-2r})=o(n^{-\frac{1}{2}}),
\end{multline*}
where the last step holds because $r>\frac{1}{4}$. Moreover,
\begin{multline*}
\E \left\{ \left|\1^{*}_{XZ}(x,1) \hat{f}_{XZ}(X,0)( W- \widehat W)\left[ \1(\widehat W \leq w)- \1( W \leq w)\right] \times  \1 (|W-w|\leq  n^{-r})\right| \right\}\\
\leq \E \left\{ \left| \hat{f}_{XZ}(X_1,0) \times (W-\widehat W)\times  \1 (|W-w|\leq  n^{-r})\right| \right\} = O(1)\times O(n^{-\iota})\times O(n^{-r})=o(n^{-\frac{1}{2}}).
\end{multline*}
Then, we have $\mathrm T_1 = o_p(n^{-\frac{1}{2}})$.

For term $ \mathrm T_2$, note that
\begin{multline*}
\E [ | \mathrm T_2 | ] \leq \frac{\overline K}{h}\times \sqrt {\E [(w-\widehat W)^2]}\times \sqrt {\P \left(|\widehat W-W|>n^{-r}\right)}\\
 \leq \frac{\overline K}{h}\times \sqrt{\E[ \widehat W^2]-2w\cdot \E [\widehat W]+ w^2} \times \sqrt {\P \left[|\hat \delta(X)-\delta(X)|>n^{-r}\right]},
\end{multline*}
where $\overline K$ is the upper bound of $K(\cdot)$. Because $W$ is a bounded random variable and $w$ belongs to a compact set, then $ \sqrt{\E[\widehat W^2] -2w\cdot \E [ \widehat W ]+ w^2}=O(1)$. 
Moreover, by Lemma \ref{lem:8}, 
$\E | \mathrm T_2|\leq o(n^{-k})$ for any $k>0$. Hence, $ \mathrm{T}_2 = o_p(n^{-\frac{1}{2}})$.

\subsection{Proof of Theorem \ref{thm:6}}
\textbf{Proof:}
By Lemma \ref{lem:5}, we have
\[
\widehat {\mathcal T}^c_n=\sqrt{n} \left| \widetilde G(w, x, 1) - \widetilde G(w, x, 0) \right|+o_p(1).
\] 
Note that
\[
\widetilde G(w,x,z) = \mathrm U_{1}(w,x,z)+\mathrm U_{2}(w,x,z)+o_p(n^{-1/2})
\]
where
\begin{align*}
&\mathrm U_{1}(w,x,z)\equiv  \frac{1}{n}\sum_{i=1}^n \1^*_{W_i}(w) \times \1^{*}_{X_iZ_i}(x,z) \times  \hat{f}_{XZ}(X_i,z')\times(W_i-\widehat W_i);\\
&\mathrm U_{2}(w,x,z)\equiv \frac{1}{n}\sum_{i=1}^n \1^*_{W_i}(w) \times \1^{*}_{X_iZ_i}(x,z) \times  \hat {f}_{XZ}(X_i,z')\times(w-W_i) .
\end{align*} Therefore,
\begin{align*}
&\sqrt{n} \left[ \widetilde G(w, x, 1) - \widetilde G(w, x, 0) \right]\\
&\quad=\sqrt{n} \left\{ \mathrm U_1(w, x, 1) - \mathrm U_1(w, x, 0)-\left[\E \mathrm U_1(w,x,1)-\E \mathrm U_1(w,x,0)\right]\right\}\\
&\qquad+\sqrt{n} \left\{ \mathrm U_2(w, x, 1) - \mathrm U_2(w, x, 0)-\left[\E \mathrm U_2(w,x,1)-\E \mathrm U_2(w,x,0)\right]\right\}\\
&\qquad+\sqrt n\left[\E \mathrm U_1(w,x,1)-\E \mathrm U_1(w,x,0)\right]+\sqrt n\left[\E \mathrm U_2(w,x,1)-\E \mathrm U_2(w,x,0)\right].
\end{align*}

We first discuss  $\mathrm U_2$ terms.  
By definition,
\begin{eqnarray*}
\mathrm U_2(w,x,z)
&=&\frac{1}{n(n-1)}\sum_{i=1}^n\sum_{j\neq i}   \{  \1^{*}_{X_iZ_i}(x,z) \lambda(W_i-w) \times K_{X_j,h}(X_i) \1(Z_j=z') \} \\
&=&  \frac{1}{n(n-1)}\sum_{i=1}^n\sum_{j\neq i} \zeta_{n,ij}(w,x,z)
\end{eqnarray*}
where $\zeta_{n,ij}(w,x,z)=\1^{*}_{X_iZ_i}(x,z) \times \lambda(W_i-w) \times K_{X_j,h}(X_i) \times \1(Z_j=z')$.

Let $ \zeta^*_{n,ij}(w,x,z)=\frac{1}{2}\left[ \zeta_{n,ij}(w,x,z)+ \zeta_{n,ji}(w,x,z)\right]$.
Then, $ \zeta^*_{n,ij}$ is symmetric in indices $i$ and $j$.
Therefore,
\[
\mathrm U_{2}(w,x,z)= \frac{1}{n(n-1)}\sum_{i=1}^n\sum_{j\neq i}\zeta^*_{n,ij}(w,x,z),
\]
which is a $\mathcal{U}$-process indexed by $(w,x,z_\ell)$.  By \citet[Theorem 5]{nolan1988functional} and \citet[Lemma 3.1]{powell1989semiparametric},
\begin{eqnarray*}
&&\mathrm{U}_{2}(w,x, z)-\E[\mathrm{U}_2(w,x,z)]\\
&=& \frac{2}{n}\sum_{i=1}^n\left\{ \E[\zeta^*_{n,ij}(w,x,z)|Y_i,D_i,X_i,Z_i]- \E[\zeta^*_{n,ij}(w,x,z)]\right\}+ o_p(n^{-1/2}).
\end{eqnarray*}where the $o_p(n^{-1/2})$ applies uniformly over $(w,x)$.  Note that
\begin{multline*}
\E[\zeta^*_{n,ij}(w,x,z)|Y_i,D_i,X_i,Z_i]\\
=\frac{1}{2}\Big\{\1^*_{XZ}(x,z) f_{XZ}(X,z') \lambda(W-w)
+\1^*_{XZ}(x,z') f_{XZ}(X,z) \Pi (w|X,z)\Big\}+o_p(1).
\end{multline*}
Next, we derive $\E [\zeta^*_{n,ij}(w,x,z)]$.
Let $u_1(w,x,z)=\E [\1^*_{XZ}(x,z) f_{XZ}(X,z') \lambda(W-w)]$ and \\ $u_2(w, x,z)=\E [\1^*_{XZ}(x,z') f_{XZ}(X,z) \Pi (w|X,z)]$.
Note that under $\mathcal{H}_0$
\[
u_1(w,x,z)=u_2(w,x,z)=\int \1 (X\leq x) \Pi(w|X)f_{X|Z}(X|1)f_{X|Z}(X|0)d X \times \P(Z=1) \P(Z=0),
\] invariant with $z$.  Therefore, $\E [\zeta^*_{n,ij}(w,x,z)]=\frac{1}{2}[u_1(w,x,z)+u_2(w,x,z)]$ is also invariant with $z$. Let $u^e(w,x)=\E [\zeta^*_{n,ij}(w,x,z)]$.  Moreover, by \citet[Theorem 3.1]{powell1989semiparametric},
\begin{multline*}
\frac{2}{\sqrt n}\sum_{i=1}^n\left\{ \E[\zeta^*_{n,ij}(w,x,z)|Y_i,D_i,X_i]- \E[\zeta^*_{n,ij}(w,x,z)]\right\} \\
= \E_n\left\{\1^*_{XZ}(x,z) f_{XZ}(X,z') \lambda(W-w)-u^e(w,x) \right\}\\
+\E_n\left\{\1^*_{XZ}(x,z') f_{XZ}(X,z) \Pi(w|X,z)-u^e(w,x) \right\}+ o_p(n^{-\frac{1}{2}}),
\end{multline*}
where the $o_p(n^{-1/2})$ holds uniformly over $(w,x)$. Moreover, under $\mathcal{H}_0$, there is $\Pi(w|X,z)=\E (\lambda(W-w)|X)$. Thus,
\begin{multline*}
\mathrm U_2(w, x, 1) - \mathrm U_2(w, x, 0)-\left[\E \mathrm U_2(w,x,1)-\E \mathrm U_2(w,x,0)\right] \\
= \E_n\left\{\Big[\frac{\1^*_{XZ}(x,1)}{f_{XZ}(X,1)}-\frac{\1^*_{XZ}(x,0)}{f_{XZ}(X,0)}\Big] f_{XZ}(X,0)f_{XZ}(X,1)\big[ \lambda(W-w)-\E (\lambda(W-w)|X)\big] \right\}+o_p(n^{-\frac{1}{2}}).
\end{multline*}

We now turn to $\mathrm U_1(w,x,z)$.
Note that
\[
\mathrm U_1(w,x,z) = -\frac{1}{n}\sum_{i=1}^n \left \{  \1^{*}_{W_iX_iZ_i}(w,x,z)  f_{XZ}(X_i,z') (1-D_i)  [\hat {\delta}(X_i)- \delta(X_i)] \right\} +o_p(n^{-\frac{1}{2}}),
\]
provided that  $\E\left|\left[\hat f_{XZ}(X_i,z')- f_{XZ}(X_i,z')\right]\times \left[\hat \delta(X_i) -\delta(X_i)\right]\right|=o_p(n^{-\frac{1}{2}})$ holds.
By a similar decomposition argument on $\hat \delta (X)-\delta(X)$ in Lemma \ref{lem:8}, we have
\[
\mathrm U_1(w,x,z)=-\frac{1}{n(n-1)}\sum_{i=1}^n\sum_{j\neq i}  \xi_{n,ij}(w,x,z)+o_p(n^{-1/2})
\]
where  $\xi_{n,ij}(w,x,z)= \1^{*}_{W_iX_iZ_i}(w,x,z)  f_{XZ}(X_i,z') (1-D_i)   \frac{[W_j-\E (W_j|X_i)] K_{X_j,h}(X_i) }{p(X_i,1)-p(X_i,0)}\left[\frac{ \1(Z_j=1)}{f_{XZ}(X_i,1)} -\frac{ \1(Z_j=0)}{f_{XZ}(X_i,0)}\right]$. 
Moreover,  let $ \xi^*_{n,ij}(w,x,z)=\frac{1}{2}[ \xi_{n,ij}(w,x,z)+ \xi_{n,ji}(w,x,z)]$.   

By a similar argument for $\mathrm U_2$,
\begin{multline*}
\mathrm{U}_{1}(w,x,z) - \E [\mathrm{U}_{1}(w,x,z)]\\
= -\frac{2}{n}\sum_{i=1}^n\left\{ \E[\xi^*_{n,ij}(w,x,z)|Y_i,D_i,X_i,Z_i]- \E[\xi^*_{n,ij}(w,x,z)]\right\}+ o_p(n^{-1/2}).
\end{multline*}
Note that $\E [\xi_{n,ij}(w,x,z)|Y_i,D_i,X_i,Z_i]=0$ and
\begin{align*}
&\E [\xi_{n,ji}(w,x,z)|Y_i,D_i,X_i,Z_i] =\E \left\{\E [\xi_{n,ji}(w,x,z)|X_j,Z_j,Y_i,D_i,X_i,Z_i]\big|Y_i,D_i,X_i,Z_i\right\}\\
&= \E \Bigg\{ \1^{*}_{X_jZ_j}(x,z)  f_{XZ}(X_j,z') \P(W\leq w;D=0|X_j,Z_j)[W_i-\E(W|X_j)] \\
&\quad\times  \frac{K_{X_j,h}(X_i) }{p(X_j,1)-p(X_j,0)}\left[\frac{ \1(Z_i=1)}{f_{XZ}(X_j,1)}-\frac{ \1(Z_i=0)}{f_{XZ}(X_j,0)}\right]\Big| Y_i,D_i,X_i,Z_i\Bigg\}\\
&=  F^*_{WD|XZ}(w,0|X_i,z)  [W_i- \E(W|X_i)]  \frac{f_{XZ}(X_i,0)f_{XZ}(X_i,1)}{p(X_i,1)-p(X_i,0)}\left[\frac{ \1^*_{X_i,Z_i} (x,1)}{f_{XZ}(X_i,1)}-\frac{ \1^*_{X_i,Z_i} (x,0)}{f_{XZ}(X_i,0)}\right]+o_p(1)
\end{align*}
where the last step comes from \deleted[id=TC]{the}Bochner's Lemma \citep[see, e.g.,][]{rudin2017fourier} and uses the fact the integrant equals zero if $Z_j=z'$.


Thus, we have
\begin{multline*}
\mathrm{U}_{1}(w,x, z)- \E[\mathrm{U}_{1}(w,x, z)]\\
= - \E_n\Bigg\{  [W- \E (W|X)]\frac{F^*_{WD|XZ}(w,0|X,z)}{p(X,1)-p(X,0)} \left[\frac{ \1^*_{XZ}(x,1)}{f_{XZ}(X,1)}-\frac{ \1^*_{XZ}(x,0)}{f_{XZ}(X,0)}\right] f_{XZ}(X,1) f_{XZ}(X,0)\Bigg\} + o_p(n^{-\frac{1}{2}}),
\end{multline*}where the $o_p(n^{-1/2})$ holds uniformly over $(w,x)$. It follows that
\[
\mathrm U_{1}(w,x, 1)- \E \mathrm U_{1}(w,x, 1)-[\mathrm U_{1}(w,x, 0)- \E \mathrm U_{1}(w,x, 0)]= \E_n\phi^c_{wx} + o_p(n^{-\frac{1}{2}}).
\]

By Assumption 9, we have $\E[\mathrm{U}_1(w,x;z)]=o_p(n^{-\frac{1}{2}})$.  Therefore, under  $\mathcal{H}_0$,
\begin{align*}
&\sqrt{n} \left[ \widetilde G(w, x, 1) - \widetilde G(w, x, 0) \right]\\
&=\sqrt{n} \left\{ \mathrm{U}_1(w, x, 1) - \mathrm{U}_1(w, x, 0)-\left\{\E[\mathrm{U}_1(w,x,1)]-\E[\mathrm{U}_1(w,x,0)]\right\}\right\}\\
&\quad+\sqrt{n} \left\{ \mathrm U_2(w, x, 1) - \mathrm U_2(w, x, 0) - \left\{ \E[\mathrm{U}_2(w,x,1)]-\E[\mathrm{U}_2(w,x,0)]\right\}\right\} + o_p(1)\\
&= \sqrt n \times \E_n (\psi^c_{wx}+\phi^c_{wx})+ o_p(1),
\end{align*}
which converges to a zero-mean Gaussian process  with the given covariance kernel.

\subsection{Proof of Theorem \ref{thm: size-conti}}
The proof is similar to that of Theorem \ref{thm: size-discrete} and we skip it for brevity. \\
\noindent{\bf Discussion on Assumptions \ref{assu: bias-continuous}, \ref{assu: bias-strengthen} and
\ref{assu: multi-conti}:}
By \cite{masry1996multivariate} and Lemma A.1 of \cite{abrevaya2015estimating}, we have the bias terms satisfying
\begin{align*}
&\sup_{x\in\mathcal{S}_X^\xi}\Big|\E \Big[\frac{1}{n}\sum_{j=1}^n \1_{Z_j}(z) K_{X_j,h}(x)\Big]-f_{XZ}(x,z)\Big|= O(h^k),\\
&\sup_{x\in\mathcal{S}_X^\xi}\Big|\E \Big[\frac{1}{n}\sum_{j=1}^n D_j \1_{Z_j}(z) K_{X_j,h}(x) \Big]-p(x,z) f_{XZ}(x,z)\Big|= O(h^k),\\
&\sup_{x\in\mathcal{S}_X^\xi}\Big|\E\Big[\frac{1}{n}\sum_{j=1}^n Y_j \1_{Z_j}(z) K_{X_j,h}(x)\Big]-\E (Y|X=x,Z=z) f_{XZ}(x,z)\Big|= O(h^k),\\
&\sup_{x\in\mathcal{S}_X^{\xi}}\big|\E [\hat \delta(x)]-\delta(x)\big|=O(h^k),
\end{align*}
where $k$ is the order of kernel $K$.  According to Assumptions \ref{assu: bias-continuous} and \ref{assu: bias-strengthen}, we require that $h^k=o(n^{-1/2})$ and $\sqrt{nh^{d_X}}$ diverges at a rate faster than $n^{\iota}$ with $\iota>1/4$.   Therefore, if we pick a $k$-th order Epanechnikov kernel with $k=2 d_X + 2$, $h=O(n^{1/(2d_X +1)})$,  then Assumptions \ref{assu: bias-continuous} and \ref{assu: bias-strengthen} will be satisfied. 
Note that under the same conditions, Assumptions \ref{assu: multi-conti} (i), (iii) and (iv) are also satisfied.

We next consider Assumption \ref{assu: multi-conti} (ii).  To estimate $E[W|X=x]$, we will replace $W_i$'s with $\widehat{W}_i$'s in the kernel regressions.  Because $|W_i-\widehat{W}_i|=o_p(1)$ uniformly, we can show that Assumption \ref{assu: multi-conti} (ii) holds. 

We finally consider  Assumption \ref{assu: multi-conti} (iv).  Define the estimator for $F_{W|DXZ}(w|0,x,z)$ as
\begin{align*}
\widehat{F}_{W|DXZ}(w|0,x,z)=\frac{\sum_{i=1}^n \1(\widehat{W}_i\leq w) \1(D_i=0, Z_i=z) K^F_{X_i,h_F}(X_i) }{\sum_{i=1}^n \1(D_i=0, Z_i=z) K^F_{X_i,h_F}(X_i)},
\end{align*} 
where $K^F$ is a regular Epanechnikov kernel and $h_F$ is the bandwidth. 
It straightforward to see that if $h_F\rightarrow 0$ and $nh^{d_X}\rightarrow \infty$, then we can show that $\sup_{x\in\mathcal{S}^\xi_X,z\in\mathcal{S}_Z, w\in\mathcal{S}_W}|\widehat{F}_{W|DXZ}(w|0,x,z)-F_{W|DXZ}(w|0,x,z)| \stackrel{p}{\rightarrow} 0$.  Given that the regular Epanechnikov kernel is always non-negative, $\widehat{F}_{W|DXZ}$ $(w|0,x,z)$ is monotonically increasing as well.  
Hence, Assumption \ref{assu: multi-conti} (iv) also holds.

\section{Technical Lemmas}
Let  $\Delta p(x)\equiv p(x,1) - p(x,0)$, which is strictly positive by Assumption \ref{assu: 5}.

\begin{lem}\label{lem:7}
Suppose Assumptions 1 and 5 hold. Then, we have
\begin{multline}
\label{delta_influence}
\sqrt n[\hat \delta(x)-\delta(x)]= \frac{1}{\Delta p(x)}\times \sqrt n\E_n \left\{\big[W -\E (W|X=x,Z=0)\big]\times \frac{\1_{XZ}(x,1)}{\P (X=x,Z=1)} \right\} \\
-\frac{1}{\Delta p(x)}\times \sqrt n \E_n \left\{\big[W -\E (W|X=x,Z=1)\big]\times\frac{\1_{XZ}(x,0)}{\P(X=x,Z=0)}\right\}+ o_p(1).
\end{multline}
\end{lem}


\noindent
\textbf{Proof of Lemma \ref{lem:7}:}
Fix $X=x$. For expositional simplicity, we suppress $x$ in the following proof. Moreover, let $\mathrm A _{n}(z)=\E_n [Y \1_{XZ}(x,z) ]$, $\mathrm B_{n}(z) =\E_n [D \1_{XZ}(x,z)]$, $\mathrm C_{n}(z) =\E_n  \1_{XZ}(x,z)$, $\mathrm A (z)= \E [Y \1_{XZ}(x,z)]$, $\mathrm B (z)=\E [D \mathrm 1_{XZ}(x,z)]$ and $ \mathrm C (z)= \E  \1_{XZ}(x,z)= \P (X=x,Z=z)$.  By definition, note that
\[
\hat \delta(x)=\frac{\mathrm A_{n}(1) \mathrm C_n(0)-\mathrm A_{n}(0)\mathrm C_n(1)}{\mathrm B_n(1) \mathrm C_n(0)-\mathrm B_n(0)\mathrm C_n(1)} \ \ \text{ and } \  \delta(x)=\frac{\mathrm A(1) \mathrm C(0)-\mathrm A(0)\mathrm C(1)}{\mathrm B(1) \mathrm C(0)-\mathrm B(0)\mathrm C(1)}.
\]  It follows that
\begin{multline*}
\hat \delta(x)-\delta(x)=\frac{\mathrm A_{n}(1) \mathrm C_n(0)-\mathrm A_{n}(0)\mathrm C_n(1)-[\mathrm A(1) \mathrm C(0)-\mathrm A(0)\mathrm C(1)]}{\mathrm B_n(1) \mathrm C_n(0)-\mathrm B_n(0)\mathrm C_n(1)}\\
+\left\{ \frac{\mathrm A(1) \mathrm C(0)-\mathrm A(0)\mathrm C(1)}{\mathrm B_n(1) \mathrm C_n(0)-\mathrm B_n(0)\mathrm C_n(1)}-\frac{\mathrm A(1) \mathrm C(0)-\mathrm A(0)\mathrm C(1)}{\mathrm B(1) \mathrm C(0)-\mathrm B(0)\mathrm C(1)}\right\}
\equiv \mathrm I+ \mathrm {II}.
\end{multline*}

We first look at term $\mathrm I$. By the central limit theorem and Assumption 5, we have $\mathrm A_n(z)=\mathrm A(z)+O_p(n^{-1/2})$, $\mathrm B_n(z)=\mathrm B(z)+O_p(n^{-1/2})$ and $\mathrm C_n(z)=\mathrm C(z)+O_p(n^{-1/2})$. Therefore,
\begin{align*}
\mathrm I &=\frac{\left[\mathrm A_{n}(1)-\mathrm A(1)\right]  \mathrm C(0) + \mathrm A(1) \left[\mathrm C_n(0)-\mathrm C(0)\right]}{\mathrm B(1) \mathrm C(0)-\mathrm B(0)\mathrm C(1)}\\
&\quad- \frac{\left[\mathrm A_{n}(0)-\mathrm A(0)\right] \mathrm C(1)+\mathrm A(0) \left[ \mathrm C_n(1)-\mathrm C(1)\right]}{\mathrm B(1) \mathrm C(0)-\mathrm B(0)\mathrm C(1)}+o_p(n^{-1/2})\\
&=\frac{\mathrm A_{n}(1)  \mathrm C(0) -\mathrm A(0) \mathrm C_n(1)-\mathrm A_{n}(0)  \mathrm C(1) + \mathrm A(1) \mathrm C_n(0)}{\mathrm B(1) \mathrm C(0)-\mathrm B(0)\mathrm C(1)}\\
&\quad+ \frac{2\left[\mathrm A(0)\mathrm C(1)-\mathrm A(1)\mathrm C(0)\right]}{\mathrm B(1) \mathrm C(0)-\mathrm B(0)\mathrm C(1)}+o_p(n^{-1/2}).
\end{align*}Specifically, we have
\begin{align*}
\mathrm I &= \E_n \left\{\big[Y -\E (Y|X=x,Z=0)\big] \times \1_{XZ}(x,1) \right\}\times\frac{ \P(X=x,Z=0)}{\mathrm B(1) \mathrm C(0)-\mathrm B(0)\mathrm C(1)}\\
&\quad-\E_n \left\{\big[Y -\E (Y|X=x,Z=1)\big] \times \1_{XZ}(x,0) \right\}\times  \frac{\P(X=x,Z=1)}{\mathrm B(1) \mathrm C(0)-\mathrm B(0)\mathrm C(1)}\\
&\quad+\frac{2\left[\mathrm A(0) \mathrm C(1)-\mathrm A(1)  \mathrm C(0)\right] }{\mathrm B(1) \mathrm C(0)-\mathrm B(0)\mathrm C(1)}+o_p(n^{-1/2})\\
&= \frac{1}{\Delta p(x)}\times \E_n \left\{\big[Y -\E (Y|X=x,Z=0)\big]\times \frac{\1_{XZ}(x,1)}{\P (X=x,Z=1)} \right\} \\
&\quad-\frac{1}{\Delta p(x)}\times \E_n \left\{\big[Y -\E (Y|X=x,Z=1)\big]\times\frac{\1_{XZ}(x,0)}{\P(X=x,Z=0)}\right\}- 2\delta(x) + o_p(n^{-1/2}).
\end{align*}

For term $\mathrm I\mathrm I$, by a similar argument we have
\begin{multline*}
\mathrm {II}
=\frac{-\delta(x)}{\Delta p(x)}\times \E_n \left\{\big[D -p(x,0)\big]\times \frac{\1_{XZ}(x,1)}{ \P(X=x,Z=1)}\right\} \\
+\frac{\delta(x)}{\Delta p(x)}\times \E_n \left\{\big[D -p(x,1)\big]\times \frac{\1_{XZ}(x,0)}{\mathrm P(X=x,Z=0)}\right\} + 2  \delta(x) +o_p(n^{-1/2}).
\end{multline*}
By definition of $W$, we have $W-\E [W|X=x,Z=z]=Y-\E [Y|X=x,Z=z]- [D-p(x,z)]\times \delta(x)$. 
Summing up $\mathrm I$ and $\mathrm I \mathrm I$,  we  obtain \eqref{delta_influence}.

\begin{lem}\label{lem:8}
Suppose Assumptions 8-10 hold.
Then for any $k>0$ and $r\in(\frac{1}{4},\iota)$,
\[
\sup_{x\in\mathcal S_X}n^k \times \P \left[|\hat\delta(x)-\delta(x)|> n^{-r}\right]\rightarrow 0.
\]
\end{lem}


\noindent
\textbf{Proof of Lemma \ref{lem:8}:}
First, by a similar decomposition of $\hat \delta(x)-\delta(x)$ as that in the proof of Lemma \ref{lem:7}, it suffices to show
\begin{align*}
&\sup_x n^k \times \P\left\{\left|a_{n}(x,z) - a(x,z)\right|>\lambda_{a}\times n^{-r}\right\}\rightarrow 0;\\
&\sup_xn^k \times \P\left\{\left|b_{n}(x,z) - b(x,z)\right|>\lambda_{b}\times n^{-r}\right\}\rightarrow 0;\\
&\sup_xn^k \times \P\left\{\left|q_{n}(x,z) - q(x,z)\right|>\lambda_{q}\times n^{-r}\right\}\rightarrow 0,
\end{align*}
where  $\lambda_{a}$, $\lambda_{b}$ and $\lambda_{q}$ are strictly positive constants, and
\begin{align*}
&a_{n}(x,z)= \frac{1}{n}\sum_{j=1}^n Y_j K_{X_j,h}(x) \1(Z_j=z),  \ \ \ a(x,z)=\E(Y|X=x,Z=z) \times q(x,z);\\
&b_{n}(x,z)= \frac{1}{n}\sum_{j=1}^n D_j K_{X_j,h}(x) \1(Z_j=z), \ \ \ b(x,z)=\E(D|X=x,Z=z) \times q(x,z);\\
&q_{n}(x,z)= \frac{1}{n}\sum_{j=1}^n K_{X_j,h}(x) \1(Z_j=z).
\end{align*}
For expositional simplicity, we only show the first result. It is straightforward that the rest follow a similar argument.

Let $T_{nxzj}=Y_jK(\frac{X_j-x}{h})\mathrm 1(Z_j=z)$ and 
$\tau_{nxz}= h\times \left[\lambda_{a}  n^{-r}-|\E [a_{n}(x,z)]-a(x,z)|\right]$.  Note that
\begin{align*}
&\P \left[|a_{n}(x,z)-a(x,z)|> \lambda_{a} \times n^{-r}\right]\\
&\leq \P \left[| a_{n}(x,z)-\E [a_{n}(x,z)]|+|\E [a_{n}(x,z)]-a(x,z)|> \lambda_{a} \times n^{-r}\right]\\
&\quad=\P \left\{\frac{1}{n}\left|\sum_{j=1}^n\left( T_{nxzj}-\E [T_{nxzj}]\right)\right|> \tau_{nxz}\right\}.
\end{align*}
Moreover, by Bernstein's tail inequality,
\[
\P \left\{\frac{1}{n}\left|\sum_{i=1}^n \left( T_{xzj}- \E[ T_{xzj}] \right)\right|> \tau_{nxz}\right\}
\leq 2 \exp\left(-\frac{n \times \tau_{nxz}^2}{2\text{Var}\left( T_{nxzj} \right)+\frac{2}{3}\overline K\times \tau_{nxz}}\right).
\]
where $\overline K$ is the upper bound of kernel $K$.

By Assumption 10, $|\E [a_{n}(x,z)]-a(x,z)|=O(n^{-\iota})=o(n^{-r})$. Then, for sufficiently large $n$, there is $ 0.5\lambda_{a} n^{-r}h\leq \tau_n(x,z)\leq \lambda_{a} n^{-r}h$. Moreover,
\[
\text{Var} \left( T_{nxzj} \right) \leq \E[ T^2_{nxzj}] \leq  \E \big[\E (Y^2|X)K^2(\frac{X-x}{h})\big]\leq Ch,
\]
where $C=\sup_{x} \E [Y^2|X=x]\times \sup_x f_X (x)\times \overline K \times  \int |K(u)| du<\infty$. It follows that
\[
\P\left\{\frac{1}{n}\left|\sum_{\ell=1}^n\left( T_{xzj}-\E[T_{xzj}]\right)\right|>\tau_{nxz}\right\} \leq 2 \exp \left(-\frac{\frac{\lambda_{a}}{4}nhn^{-2r}}{2C +\frac{2}{3}\overline K \lambda_{a} n^{-r}}\right).
\]
For sufficiently large $n$, we have $\frac{2}{3}\overline K \lambda_{a} n^{-r}\leq 1$.  Therefore,  for sufficiently large $n$,
\[
\P\left\{\frac{1}{n}\Big|\sum_{\ell=1}^n\left(T_{xzj}-\E[ T_{xzj}]\right)\Big|>\tau_{nxz}\right\} \leq 2 \exp\left(-\frac{  n^{2\iota-2r}}{2C +1}\right) =o(n^{-k})
\]where the inequality comes from Assumption 10.  Note that the upper bound does not depend on $x$ or $z$. Therefore,
\[
\sup_{x,z}\P \left[|a_{n}(x,z)-a(x,z)|> \lambda_{a} \times n^{-r}\right]=o(n^{-k}).
\]

\section{Testing with both discrete and continuous covariates}\label{sec: discrete-conti}
We briefly discuss how to implement our test when the covariates contain both discrete and continuous variables.  Let $X=(X_d', X_c')$ where $X_d$ is a $d_{X_d}$-dimensional vector of discrete variables taking a finite number of values in $\mathcal{X}_d$, and $X_c$ is a $d_{X_c}$-dimensional  vector of continuous covariates with support  $\mathcal{X}_c=\prod_{j=1}^{d_{X_c}}[x_{c,\ell j}, x_{c,uj}]$. Define  $\mathcal{X}^\xi_c=\prod_{j=1}^{d_{X_c}}[x_{c,\ell j}+\xi, x_{c,uj}-\xi]$

Note that in this case, $W \perp \!\!\!\perp Z | X$ is equivalent to 
\begin{align}
\Pi(w|x_c,x_d,0)=\Pi(w|x_c,x_d,1),~\forall(w,x_c,x_d)\in \mathcal{S}_{WX_cX_d}. \label{eq: multi 1}
\end{align}
With a similar argument, we can further show that (\ref{eq: multi 1}) is equivalent to
\begin{align}
&G(w,x_c,x_d,0)=G(w,x_c,x_d,1),~\forall(w,x_c,x_d)\in \mathcal{S}_{WX_cX_d},\label{eq: multi 2}
\end{align}
where for $z=0,1$,
\begin{align*}
G(w,x_c,x_d,z)=\E[\lambda(W-w) \1^{*}_{X_c}(x_c)\1_{X_dZ}(x_d,z) f_{X_c|X_dZ}(x_c|x_d,z')\Pr(X_d=x_d,Z=z')]
\end{align*}
To see this, by the same arguments as in the previous subsection, we have
\begin{align*}
G(w,x_c,x_d,z)=&\E \left[\lambda (W-w) \1^*_{X_c}(x_c) f_{X_c|X_dZ}(X_c|x_d,z') |X_d=x_d,Z=z\right] \\
&~\cdot \P(X_d=x_d,Z=0) \P(X_d=x_d,Z=1).
\end{align*}


We estimate $\delta(X_{di},X_{ci})$ by $\hat{\delta}(X_{di},X_{ci} = \hat{\delta}_1(X_{di},X_{ci})/ \hat{\delta}_2(X_{di},X_{ci})$, where
\begin{align*}
\hat{\delta}_1(X_{di},X_{ci}) & = \hbox{$\sum_{j\neq i,X_{dj}=X_{di}}$} Y_j Z_j K_{X_{cj},h}(X_{ci}) \times \hbox{$\sum_{j\neq i,X_{dj}=X_{di}}$} K_{X_{cj},h}(X_{ci}) \\
&\quad- \hbox{$\sum_{j\neq i,X_{dj}=X_{di}}$} Y_j  K_{X_{cj},h}(X_{ci}) \times \hbox{$\sum_{j\neq i,X_{dj}=X_{di}}$} Z_j K_{X_{cj},h}(X_{ci}) \\
\hat{\delta}_1(X_{di},X_{ci}) &= \hbox{$\sum_{j\neq i,X_{dj}=X_{di}}$} D_j  Z_j K_{X_{cj},h}(X_{ci}) \times \hbox{$\sum_{j\neq i,X_{dj}=X_{di}}$} K_{X_{cj},h}(X_{ci}) \\
&\quad- \hbox{$\sum_{j\neq i,X_{dj}=X_{di}}$} D_j  K_{X_{cj},h}(X_{ci}) \times \hbox{$\sum_{j\neq i,X_{dj}=X_{di}}$} Z_j K_{X_{cj},h}(X_{ci})
\end{align*}

Moreover, let $f_{X_dX_cZ}(x_c,x_d,z')=f_{X_c|X_dZ}(x|x_d,z) \P(X_d=x_d,Z=z)$, which can be estimated by
\begin{align*}
&\hat{f}_{X_cX_dZ}(X_{ci}, X_{di} ,z) = \frac{1}{n} \sum_{j\neq i,X_{dj}=X_{di}} K_{X_{cj},h}(X_{ci}) \1_{Z_j}(z).
\end{align*}
In turn, we let $\widehat W_i = Y_i + (1-D_i)\times  \hat\delta(X_{ci},X_{di})$ and can estimate ${G}(w,x_c,x_d, z)$ as 
\begin{align*}
&\widehat{G}(w,x_c,x_d, z) = \frac{1}{n}
\sum_{\{i:X_{ci}\in\mathcal{X}_c^\xi\}} \lambda (\widehat{W}_i-w) 
\1^{*}_{X_c}(x_c)\1_{X_dZ}(x_d,z)  \hat{f}_{X_cX_dZ}(X_{ci}, X_{di} ,z').
\end{align*} 
and define our test statistic as follows:
\begin{align*}
\widehat{\mathcal{T}}^{m}_{n} = 
\sup_{w\in\mathcal{S}_W,x_c\in\mathcal{X}^{\xi}_c,x_d\in\mathcal{X}_d}\ \sqrt{n} \left| \widehat{G}(w, x_c,x_d, 0) - \widehat{G}(w, x_c,x_d,1) \right|.
\end{align*}

Here, we provide the influence functions for $ \sqrt{n}(\widehat{G}(w, x_c,x_d, 0) - \widehat{G}(w, x_c, $ $x_d,1)-{G}(w, x_c,x_d, 0) - {G}(w, x_c,x_d,1)).$
Let $F^*_{WD|X_c X_d Z}(w,d|x_c, x_d,z) \equiv  F_{W|D X_C X_d Z}(w|d,x_c, x_d,z) \times \P (D=d|X_c=x_c,X_d=x_d,Z=z)$ and
\[
\kappa^m(w,x_c, x_d)= -\frac{F^*_{WD|X_c X_d Z}(w,0|x_c, x_d,1)-F^*_{WD|X_c X_d Z}(w, 0|x_c, x_d, 0)}{p(x_c,x_d,1)-p(x_c,x_d,0)}.
\]
Moreover, we define 
\begin{align*}
\psi^m_{wx} &= \1(X\in\mathcal{S}_{X_c}^\xi) \Bigg\{ \Big[\lambda(W-w) - \E[\lambda(W-w)|X_c, X_d=x_d, Z=1]\Big] \times \frac{\1^{*}_{X_c}(x_c)\1_{X_dZ}(x_d,0)}{f_{X_cX_dZ}(X_c,x_d,0)} \\
&\qquad- \Big[\lambda(W-w) - \E[\lambda(W-w)|X_c, X_d=x_d, Z=0] \Big] \times \frac{ \1^{*}_{X_c}(x_c)\1_{X_dZ}(x_d,1)} { f_{X_c X_d Z}(X_c,x_d,1)} \Bigg\}\\
&\quad\times  f_{X_c X_d Z}(X_c,x_d,0) f_{X_c X_d Z}(X_c, x_d,1);\\
\phi^m_{wx} &= \1(X\in\mathcal{S}_{X_c}^\xi) \kappa^c(w,X_c,x_d) \Bigg\{ \Big[W - \E[W|X_c, X_d=x_d, Z=0]\Big] \times \frac{\1^{*}_{X_c}(x_c)\1_{X_dZ}(x_d,1)}{f_{X_cX_dZ}(X_c,x_d,1)} \\
&\qquad- \Big[W - \E[W|X_c, X_d=x_d, Z=1] \Big] \times \frac{ \1^{*}_{X_c}(x_c)\1_{X_dZ}(x_d,0)} { f_{X_c X_d Z}(X_c,x_d,0)} \Bigg\}\\
&\quad\times  f_{X_c X_d Z}(X_c,x_d,0) f_{X_c X_d Z}(X_c, x_d,1).
\end{align*}
Given such  influence function representations, we can implement our test as before, so we omit the details.

\newpage
\bibliography{additivity}

\bibliographystyle{econometrica}

\newpage

\begin{table}[!h]
\caption{\small Rejection probabilities ($\alpha=5\%$) in the discrete-covariates case.}  \label{table:tab1}
  \begin{center}
  \begin{tabular}{l  c  ccc ccccc  }
  \hline\hline
  $p$  &$n$    & $c=1.7$ & $c=2$ & $c=2.34$ & $c=2.6$& \\
  \hline
  \multicolumn{9}{c}{\bf Panel A: at null hypothesis with $\boldsymbol{\gamma =1}$ }  \\
                            & $1000$  & $0.0040$ & $0.0060$ & $0.0150$ & $0.0150$ \\ 
  \multirow{-1}{*}{$0.25$}  & $2000$  & $0.0190$ & $0.0280$ & $0.0270$ & $0.0460$ \\ 
                            & $4000$  & $0.0420$ & $0.0360$ & $0.0440$ & $0.0470$ \\ 
                            & $1000$  & $0.0140$ & $0.0210$ & $0.0200$ & $0.0350$ \\ 
  \multirow{-1}{*}{$0.5$}   & $2000$  & $0.0260$ & $0.0330$ & $0.0500$ & $0.0410$ \\ 
                            & $4000$  & $0.0300$ & $0.0560$ & $0.0430$ & $0.0630$ \\ 
                            & $1000$  & $0.0050$ & $0.0130$ & $0.0150$ & $0.0310$ \\ 
  \multirow{-1}{*}{$0.75$}  & $2000$  & $0.0220$ & $0.0180$ & $0.0290$ & $0.0260$ \\ 
                            & $4000$  & $0.0390$ & $0.0350$ & $0.0420$ & $0.0430$ \\ 
  \hline
  \multicolumn{9}{c}{\bf Panel B: at alternative hypothesis with $\boldsymbol{\gamma =0.75}$}  \\
                            & $1000$  & $0.0230$ & $0.0290$ & $0.0480$ & $0.0410$ \\ 
  \multirow{-1}{*}{$0.25$}  & $2000$  & $0.2070$ & $0.2750$ & $0.3070$ & $0.3300$ \\ 
                            & $4000$  & $0.7720$ & $0.7980$ & $0.8030$ & $0.8190$ \\ 
                            & $1000$  & $0.0700$ & $0.1140$ & $0.1550$ & $0.1620$ \\ 
  \multirow{-1}{*}{$0.5$}   & $2000$  & $0.4740$ & $0.5350$ & $0.5590$ & $0.5870$ \\ 
                            & $4000$  & $0.9470$ & $0.9500$ & $0.9670$ & $0.9520$ \\ 
                            & $1000$  & $0.0480$ & $0.0640$ & $0.1140$ & $0.1160$ \\ 
  \multirow{-1}{*}{$0.75$}  & $2000$  & $0.2580$ & $0.3490$ & $0.4020$ & $0.4040$ \\ 
                            & $4000$  & $0.8080$ & $0.8340$ & $0.8600$ & $0.8670$ \\
  \hline
  \multicolumn{9}{c}{\bf Panel C: at alternative hypothesis with $\boldsymbol{\gamma =0.50}$}  \\
                            & $1000$  & $0.2560$ & $0.3750$ & $0.4900$ & $0.6010$ \\ 
  \multirow{-1}{*}{$0.25$}  & $2000$  & $0.9620$ & $0.9880$ & $0.9950$ & $0.9970$ \\ 
                            & $4000$  & $1.0000$ & $1.0000$ & $1.0000$ & $1.0000$ \\ 
                            & $1000$  & $0.8010$ & $0.8840$ & $0.9470$ & $0.9620$ \\ 
  \multirow{-1}{*}{$0.5$}   & $2000$  & $1.0000$ & $1.0000$ & $1.0000$ & $1.0000$ \\ 
                            & $4000$  & $1.0000$ & $1.0000$ & $1.0000$ & $1.0000$ \\ 
                            & $1000$  & $0.5100$ & $0.6890$ & $0.8010$ & $0.8600$ \\ 
  \multirow{-1}{*}{$0.75$}  & $2000$  & $0.9990$ & $0.9990$ & $1.0000$ & $1.0000$ \\ 
                            & $4000$  & $1.0000$ & $1.0000$ & $1.0000$ & $1.0000$ \\  
  \hline
  \end{tabular}
  \end{center}
\end{table}

\begin{table}[!h]
\caption{\small Rejection probabilities ($\alpha=5\%$) in the continuous-covariates case.}  \label{table:tab2}
  \begin{center}
  \begin{tabular}{l  c  ccc ccccc  }
  \hline\hline
  $p$  &$n$    & $c=1.7$ & $c=2$ & $c=2.34$ & $c=2.6$& \\
  \hline
  \multicolumn{9}{c}{\bf Panel A: at null hypothesis with $\boldsymbol{\gamma =1}$ }  \\
                            & $1000$   & $0.0140$ & $0.0500$ & $0.0580$ & $0.0940$ \\
  \multirow{-1}{*}{$0.25$}  & $2000$   & $0.0940$ & $0.1180$ & $0.1120$ & $0.1200$ \\
                            & $4000$   & $0.1400$ & $0.1240$ & $0.0940$ & $0.0720$ \\
                            & $1000$   & $0.0320$ & $0.0280$ & $0.0360$ & $0.0620$ \\
  \multirow{-1}{*}{$0.5$}   & $2000$   & $0.0048$ & $0.0440$ & $0.0620$ & $0.0360$ \\
                            & $4000$   & $0.0460$ & $0.0380$ & $0.0660$ & $0.0440$ \\
                            & $1000$   & $0.0100$ & $0.0060$ & $0.0060$ & $0.0260$ \\
  \multirow{-1}{*}{$0.75$}  & $2000$   & $0.0340$ & $0.0300$ & $0.0200$ & $0.0220$ \\
                            & $4000$   & $0.0700$ & $0.0580$ & $0.0320$ & $0.0480$ \\
  \hline
  \multicolumn{9}{c}{\bf Panel B: at alternative hypothesis with $\boldsymbol{\gamma =0.75}$}  \\
                            & $1000$   & $0.0420$ & $0.0420$ & $0.1060$ & $0.1120$ \\
  \multirow{-1}{*}{$0.25$}  & $2000$   & $0.1520$ & $0.2980$ & $0.3540$ & $0.4380$ \\
                            & $4000$   & $0.7420$ & $0.8220$ & $0.8700$ & $0.8620$ \\
                            & $1000$   & $0.0220$ & $0.0440$ & $0.0880$ & $0.1060$ \\
  \multirow{-1}{*}{$0.5$}   & $2000$   & $0.2960$ & $0.3820$ & $0.4740$ & $0.5320$ \\
                            & $4000$   & $0.8780$ & $0.9100$ & $0.9480$ & $0.9460$ \\
                            & $1000$   & $0.0020$ & $0.0000$ & $0.0140$ & $0.0120$ \\
  \multirow{-1}{*}{$0.75$}  & $2000$   & $0.0240$ & $0.0000$ & $0.0920$ & $0.1180$ \\
                            & $4000$   & $0.3980$ & $0.5260$ & $0.5980$ & $0.7000$ \\
  \hline
  \multicolumn{9}{c}{\bf Panel C: at alternative hypothesis with $\boldsymbol{\gamma =0.50}$}  \\
                            & $1000$   & $0.1780$ & $0.3680$ & $0.5980$ & $0.7280$ \\
  \multirow{-1}{*}{$0.25$}  & $2000$   & $0.8360$ & $0.9480$ & $0.9860$ & $0.9960$ \\
                            & $4000$   & $1.0000$ & $1.0000$ & $0.9980$ & $1.0000$ \\
                            & $1000$   & $0.8500$ & $0.8460$ & $0.9780$ & $0.9940$ \\
  \multirow{-1}{*}{$0.5$}   & $2000$   & $1.0000$ & $1.0000$ & $1.0000$ & $1.0000$ \\
                            & $4000$   & $1.0000$ & $1.0000$ & $1.0000$ & $1.0000$ \\
                            & $1000$   & $0.2660$ & $0.5300$ & $0.6900$ & $0.8240$ \\
  \multirow{-1}{*}{$0.75$}  & $2000$   & $0.9960$ & $1.0000$ & $1.0000$ & $1.0000$ \\
                            & $4000$   & $1.0000$ & $1.0000$ & $1.0000$ & $1.0000$ \\ 
  \hline
  \end{tabular}
  \end{center}
\end{table}

\begin{table}[!h]
\caption{Descriptive Statistics for the National JTPA Study}
\label{table:jtpa}
\begin{center}
\begin{tabular}{lccc}
& \multirow{2}{*}{All} & $Z=1$ & $Z=0$ \\
&     & (eligible) & (not eligible) \\
\hline
Men\\
\hspace{4pt} Number of observations  & $5,102$   & $3,399$   & $1,703$ \\
\hspace{4pt} Training ($D=1$)        & $41.87\%$  & $62.28\%$   & $1.12\%$ \\
\hspace{4pt} High school or GED      & $69.32\%$  & $69.26\%$  & $69.43\%$ \\
\hspace{4pt} Married                 & $35.26\%$  & $36.01\%$  & $33.75\%$ \\
\hspace{4pt} Minorities              & $38.38\%$  & $38.69\%$  & $37.76\%$ \\
\hspace{4pt} Worked less than 13 weeks in the past year  & $40.02\%$  & $40.28\%$ & $39.05\%$ \\
\hspace{4pt} 30 months earnings & $19,147$ & $19,520$ & $18,404$ \\
\\
Women\\
\hspace{4pt} Number of observations  & $6,102$   & $4,088$   & $2,014$ \\
\hspace{4pt} Training ($D=1$)        & $44.61\%$  & $65.73\%$  & $1.74\%$ \\
\hspace{4pt} High school or GED      & $72.06\%$  & $72.85\%$  & $70.45\%$ \\
\hspace{4pt} Married                 & $21.93\%$  & $22.48\%$  & $20.82\%$ \\
\hspace{4pt} Minorities              & $40.41\%$  & $40.58\%$  & $51.86\%$ \\
\hspace{4pt} Worked less than 13 weeks in the past year  & $51.79\%$  & $51.75\%$ & $51.86\%$ \\
\hspace{4pt} 30 months earnings & $13,029$ & $13,439$ & $12,197$ \\
\hline
\end{tabular}
\begin{tablenotes}
\item[] Note: Means are reported in this table for the National JTPA study 30 months' earnings sample. 
\end{tablenotes}
\end{center}
\end{table}

\begin{sidewaystable}
\scriptsize
\begin{threeparttable}
\caption{Descriptive Statistics for the 1999 and 2000 Censuses}
\begin{center}
\label{table:fertility}
\begin{tabular}{@{\extracolsep{7pt}}lcccccc@{}}
& \multicolumn{3}{c}{1990} & \multicolumn{3}{c}{2000} \\
 \cline{2-4}  \cline{5-7} \\
& \multirow{2}{*}{All} & $Z=1$ & $Z=0$ & \multirow{2}{*}{All} & $Z=1$ & $Z=0$ \\
&     & (twin birth) & (no twin birth) &     & (twin birth) & (no twin birth) \\
\hline
Observations          & $602,767$ & $6,524$  & $596,243$ & $573,437$ & $8,569$  & $564,868$ \\
Number of children    & $1.9276$  & $2.5318$ & $1.9209$ & $1.8833$  & $2.5196$ & $1.8734$ \\
At least two children ($D=1$) & $0.6500$  & $1.0000$ & $0.6461$ & $0.6163$  & $1.0000$ & $0.6104$ \\
\\
Mother\\
\hspace{4pt} Age in years          & $29.7894$ & $29.9530$ & $29.7876$ & $30.0562$ & $30.3943$ & $30.0510$ \\
\hspace{4pt} Years of education    & $12.9196$ & $12.9623$ & $12.9191$ & $13.1131$ & $13.2615$ & $13.1108$ \\
\hspace{4pt} Black                 & $0.0637$  & $0.0757$  & $0.0636$ & $0.0724$  & $0.0816$  & $0.07228$ \\
\hspace{4pt} Asian                 & $0.0326$  & $0.0321$  & $0.0326$ & $0.0447$  & $0.0335$  & $0.0448$ \\
\hspace{4pt} Other \replaced[id=TC]{races}{Races}           & $0.0537$  & $0.0592$  & $0.0536$ & $0.0912$  & $0.0806$  & $0.0914$ \\
\hspace{4pt} Currently at work     & $0.5781$  & $0.5444$  & $0.5785$ & $0.5629$  & $0.5132$  & $0.5637$ \\
\hspace{4pt} Usual hours per week  & $24.5660$ & $23.3537$ & $24.5795$ & $25.1400$ & $23.0491$ & $25.1723$ \\

\hspace{4pt} Wage or salary income last year & $8942$ & $8593$ & $8946$ & $14200$ & $13757$ & $14206$ \\

\\
Father \\
\hspace{4pt} Age in years          & $32.5358$ & $32.7534$ & $32.5333$ & $32.9291$ & $33.3102$ & $32.9232$ \\
\hspace{4pt} Years of education    & $13.0436$ & $13.0748$ & $13.0432$ & $13.0331$ & $13.1806$ & $13.0308$ \\
\hspace{4pt} Black                 & $0.0671$  & $0.0796$  & $0.0670$ & $0.0800$  & $0.0945$  & $0.0798$ \\
\hspace{4pt} Asian                 & $0.0291$  & $0.0263$  & $0.0292$  & $0.0402$  & $0.0318$  & $0.0403$ \\

\hspace{4pt} Other \replaced[id=TC]{races}{Races}           & $0.0488$  & $0.0529$  & $0.0488$ & $0.0919$  & $0.0802$  & $0.0921$ \\
\hspace{4pt} Currently at work     & $0.8973$  & $0.8922$  & $0.8974$ & $0.8512$  & $0.8584$  & $0.8511$ \\
\hspace{4pt} Usual hours per week  & $42.7636$ & $42.7704$ & $42.7635$ & $43.8805$ & $43.8789$ & $43.8805$ \\
\hspace{4pt} Wage or salary income last year & $27020$ & $28039$ & $27010$ & $38041$ & $41584$ & $37987$ \\
\\
Parents \\
Wages or salary income last year & $35,963$ & $36,632$ & $35,956$ & $52,241$ & $55,342$ & $52,193$\\
\hline
\end{tabular}
\begin{tablenotes}
\item[] Note: Data from the $1\%$ and $5\%$ PUMS in 1990 and 2000. Own calculations using the PUMS sample weights. The sample consists of married \replaced[id=TC]{mothers}{mother} between 21 and 35 years of age with at least one child.
\end{tablenotes}
\end{center}
\end{threeparttable}
\end{sidewaystable}

\end{document}